\documentclass[twoside]{article}
\usepackage[latin1]{inputenc}
\usepackage{t1enc}
\usepackage{a4}
\usepackage{tabularx}
\usepackage{epsf}

\def\bref{\vspace{4pt}\noindent\hangindent=10mm}

\begin{document}

\setcounter{figure}{0}
\setcounter{section}{0}
\setcounter{equation}{0}

\begin{center}
{\Large\bf
High-Velocity Clouds\\[0.2cm]
and the Local Intergalactic Medium}\\[0.7cm]

{\it Ludwig Biermann Award Lecture}\\[0.7cm]

Philipp Richter\\[0.17cm]
Argelander-Institut f\"ur Astronomie (AIfA), \\
Universit\"at Bonn, Auf dem H\"ugel 71, 53121 Bonn\\
prichter@astro.uni-bonn.de
\end{center}

\vspace{0.5cm}

\begin{abstract}
\noindent{\it
In this article I review recent observations of
the gaseous halos of galaxies and the intergalactic medium
at low redshift. In the first part I discuss distribution,
metal content, and physical properties of the Galactic intermediate-
and high-velocity clouds and the hot halo of the Milky Way. 
Recent absorption and emission measurements show that the
Galaxy's tidal interaction with the Magellanic Clouds, the
infall of low-metallicity gas, as
well as the circulation of gas as part of the galactic fountain
contribute to the observed distribution of gas in the 
halo of the Milky Way.
In the second part of this article I give a short overview
on the circumgalactic gaseous 
environment of other nearby spiral galaxies.
Multi-wavelength observations demonstrate that neutral and ionized
gaseous halos of galaxies are common, and that they 
extend deep into intergalactic space.
These studies suggest that the gaseous material around spiral galaxies 
is tightly connected to the on-going hierarchical formation and evolution of
these galaxies.
In the last part of this article I summarize 
recent quasar absorption-line measurements
of the local intergalactic medium. In accordance with cosmological
simulations, absorption-line studies in the far-ultraviolet 
indicate that both the photoionized Ly\,$\alpha$ forest and the 
shock-heated warm-hot intergalactic medium harbor
a substantial fraction of the baryons in the local Universe.
}
\end{abstract}

\section{Introduction}

Studying the distribution and physical properties of low-density gas that
is situated in the extended halos of galaxies and in the
intergalactic medium is of fundamental importance for
our understanding of the formation and evolution of 
structure in the Universe.
Multi-wavelength observations of gaseous halos and 
circumgalactic gas of galaxies provide a detailed
insight into the various complex processes that balance
the exchange of gaseous matter and energy between
individual galaxies and the intergalactic medium.
Supernova explosions in spiral galaxies
create large cavities filled with hot gas in the gaseous disk.
This gas eventually breaks out of the disk and flows into
the halo and into intergalactic space, where it enriches
the circumgalactic gas and the intergalactic medium with
heavy elements.    
In addition, the interaction and the merging of galaxies
transports large amounts of interstellar
material into the halos and immediate intergalactic environment
of galaxies. Gas that is related to tidal interactions
of galaxies therefore indicates the on-going growth and
evolution of galaxies in terms of the hierarchical 
structure-evolution in the Universe. 
Finally, knowledge about the distribution and physical properties of the 
intergalactic medium is important to understand the global
picture of how large-scale structures form and how
the baryonic matter is distributed throughout the Universe. \\
\indent With the availability of space-based spectroscopic
instruments operating in the Ultraviolet and
Far-Ultraviolet (UV and FUV, respectively), such as the
{\it Orbiting and Retrievable Far and Extreme
Ultraviolet Spectrometer} (ORFEUS), the {\it
Space Telescope Imaging Spectrograph} (STIS), and
the {\it Far Ultraviolet Spectroscopic
Explorer} (FUSE) it has become possible to
explore the gaseous halos of the Milky Way and other galaxies 
and the local intergalactic medium in absorption against distant extragalactic
UV background sources like quasars (QSOs) and
Active Galactic Nuclei (AGN). The UV and FUV spectroscopic
range is particularly interesting for studying the
low-density, multi-phase circumgalactic medium, because
many atomic and molecular species and their
ions have most of their electronic transitions in the
region between 900 and 3000 \AA\,
(e.g., H$_2$, H\,{\sc i}, D\,{\sc i}, C\,{\sc ii}, C\,{\sc iii}, C\,{\sc iv}, 
N\,{\sc i}, O\,{\sc i}, 
O\,{\sc vi}, Si\,{\sc ii}, Si\,{\sc iii}, and Fe\,{\sc ii}).
Measurements of absorption lines from these
species together with supporting emission-line observations
in other wavelength ranges 
therefore allow us to analyze in detail the gas
in the halos of galaxies and in the intergalactic 
medium in all of its phases (i.e., from molecular to 
highly-ionized).\\
\indent This article is the written version of my Ludwig Biermann
Award Lecture that was held in September 2005 at the 
annual meeting of the Astronomische Gesellschaft in
Cologne/Germany. The article is not meant to represent
an overall review on high-velocity clouds and the
local intergalactic medium, but  
rather reflects my personal view on this topic together
with an overview of my own contributions in this field. 
This article is organized as follows. In Section 2 I discuss
distribution and properties of the intermediate- and high-velocity
clouds in the halo of the Milky Way. In Section 3, I review the hot
Galactic corona and highly-ionized gas in the circumgalactic
environment of the Galaxy. Section 4 deals with circumgalactic
gaseous structures in other galaxies. In Secion 5, properties
and baryon content of the local intergalactic medium are discussed.
A short conclusion is given in Section 6.

\section{The Milky Way's High-Velocity Clouds}

\subsection{Overview}

Ever since their detection more than forty years ago, the origin
and nature of the intermediate and high-velocity 
clouds (IVCs and HVCs, respectively)
in the halo of the Milky Way has been controversial.
These gas clouds are concentrations
of neutral hydrogen (H\,{\sc i}) with radial
velocities that are inconsistent
with a simple model of differential galactic rotation. They
are thought to be located in the inner and outer halo and nearby 
intergalactic environment of the Milky Way.
The distinction between IVCs and HVCs is loosely based on the
observed radial velocities of the clouds;
IVCs have radial velocities with respect to the
Local Standard of Rest (LSR) of $30$ km\,s$^{-1} \leq |V_{\rm LSR}|
\leq 90$ km\,s$^{-1}$, typically, while
HVCs have typical velocities $|V_{\rm LSR}| > 90$ km\,s$^{-1}$.

This extraplanar neutral gas with high radial velocities was
first detected in optical absorption spectra of stars at high 
galactic latitudes (M\"unch 1952; M\"unch \& Zirin 1961).
Based on these observations, Spitzer (1956) predicted 
that the neutral halo clouds must be embedded in a Corona
of hot gas that provides the thermal pressure necessary to 
to keep the clouds together
(see Section 3).
The first detection of the IVCs and HVCs in H\,{\sc i} 21cm
radio emission was reported some years later by Muller et al.\,(1963).
Oort (1970) proposed 
that these clouds represent condensed 
gaseous relicts from the formation phase of the
Milky Way.  
This idea later was revived by Blitz et al.\,(1999), who 
suggested that HVCs represent the building blocks of galaxies in a 
hierarchical galaxy formation scenario. 
Since the Galaxy is 
surrounded by smaller satellite galaxies (e.g., the 
Magellanic Clouds), another explanation is that IVCs and HVCs
are gaseous streams related to the
merging and accretion of these satellites by the Milky Way. 
In this picture, HVCs would be the gaseous counterparts of the Milky Way's 
circumgalactic stellar streams, which are believed to 
represent the relics of dwarf galaxies that have been
accreted by the Milky Way (e.g., Ibata 1994). While all these models
assume that HVCs are truly extragalactic objects that are about to
merge with the Galaxy from outside, there are other scenarios 
that see the IVCs and HVCs as objects that have their 
origin in the disk of the Milky Way, e.g., as part of the 
so-called "galactic fountain". In the galactic fountain
model (Shapiro \& Field 1976; Houck \& Bregman 1990), hot gas is ejected
out of the Galactic disk by supernova explosions, and part of
this gas falls back in the form of condensed neutral
clouds that move at intermediate and high radial velocities.
Whatever the origin of the Milky Way's IVCs and HVCs is, it
has become clear that they must play an important role in
the evolution of our Galaxy. 

Two extremely important parameters to distinguish between
the Galactic and extragalactic models of IVCs and HVCs
are the {\it distance} and the {\it chemical 
composition} of these clouds.
Both distance and metal abundance measurements require  
the use of high-resolution absorption line spectra.
Consequently, absorption
line measurements in the FUV and in the optical
have been used extensively during the last few years to obtain 
new information about the chemical
composition and spatial distribution of IVCs and HVCs.
These studies have lead to a much improved 
understanding of the nature of 
these clouds (see also Wakker \& Richter 2004). 
In the following subsections I want to 
highlight some of these measurements.

\begin{figure}[t!]
\epsfxsize=12.5cm
\epsfbox{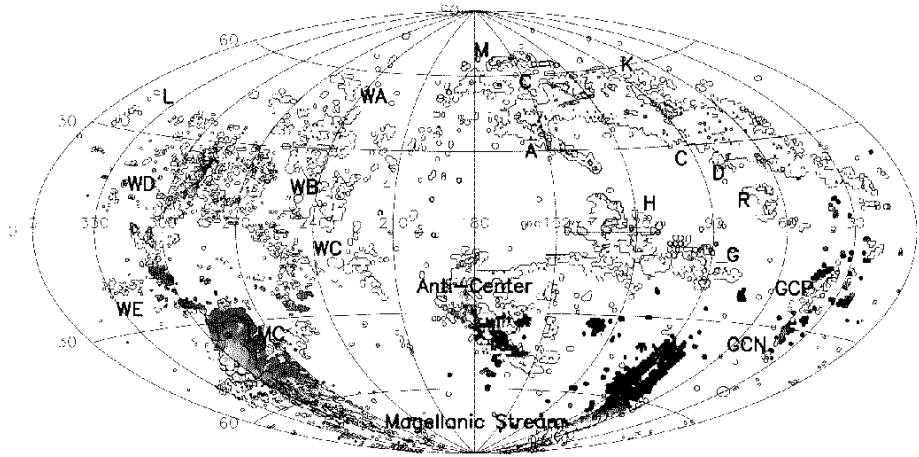}
\caption{
Aitoff projection all-sky map of the Galactic HVCs, in galactic 
coordinates, for H\,{\sc i} column densities $>2\times 10^{18}$ cm$^{-2}$,
based on the 21cm data from Hulsbosch \& Wakker (1988) and Morras et al.\,(2000).
Individual HVC complexes are indicated. Figure taken from
Wakker (2004) [PLEASE CONTACT THE AUTHOR FOR A HIGH-QUALITY VERSION OF THIS FIGURE.]
}
\end{figure}

\subsection{Distribution of Neutral Gas in the Halo}

The distribution of neutral gas in the halo of the Milky Way
in the form of IVCs and HVCs
can be studied best in the H\,{\sc i} 21cm line of
neutral hydrogen. Fig.\,1 shows the H\,{\sc i} HVC sky (Aitoff projection) for
H\,{\sc i} column densities $>2\times 10^{18}$ cm$^{-2}$ 
based on the
the 21cm data from Hulsbosch \& Wakker (1988) and Morras et al.\,(2000).
For $N$(H\,{\sc i}$)>2\times 10^{18}$ cm$^{-2}$ the sky-covering
fraction of HVC H\,{\sc i} gas is $\sim 30$ percent (see, e.g., 
Murphy et al.\,1995; Wakker 2004), while for 
$N$(H\,{\sc i}$)>7\times 10^{17}$ cm$^{-2}$ the covering fraction
increases to almost 50 percent. Note that 
neutral gas structures with even lower H\,{\sc i} column densities
do exist. Such clouds lie below the detection limit of
current 21cm radio observations, but they can be observed using
optical and FUV absorption line measurements (e.g., Richter et al.\,2005a;
see also Section 4.3).
The H\,{\sc i} HVC sky is divided into individual HVC "complexes" 
(for a list of names see Wakker 2001),
as indicated in Fig.\,1. The most prominent of these HVC complexes 
are complex A ($l\sim 150, b\sim +35$), complex C ($l\sim 30-150, b\sim +40$),
complex H ($l\sim 130, b\sim 0$), the Anti-Center Cloud ($l\sim 180, b\sim -30$),
and the Magellanic Stream ($l\sim 90-300, b\sim -30$ to $-90$).
Recently, Kalberla et al.\,(2005) have combined and newly reduced
21cm data from the Leiden-Dwingeloo Survey 
(LDS; Hartmann \& Burton 1997)
and the Villa-Elisa Survey (Morras et al.\,2000)
and have created a H\,{\sc i} 21cm all-sky survey 
(the Leiden-Argentina-Bonn survey; LAB survey) that
represents an excellent data base to study the distribution of the HVC H\,{\sc i}
gas in the Milky Way halo.\\
\indent Next to the large HVC complexes listed above 
there is a population of isolated, relatively
small HVCs, commonly referred to as compact HVCs 
(CHVCs; Braun \& Burton 1999). These are sharply 
bounded in angular extent (angular sizes are $<2^{\circ}$,
typically) and appear to have no kinematic or spatial connection to other HVC
features. It was suggested that CHVCs are  
candidates for dark-matter halos filled with gas at megaparsec 
distances that are 
distributed throughout the 
Local Group of galaxies (Braun \& Burton 2000).
However, recent distance estimates for these
objects (Westmeier et al.\,2005a) speak against this scenario.\\
\indent Also the intermediate-velocity clouds in the halo cover a significant portion of the sky.
Fig.\,2 shows the H\,{\sc i} IVC sky in the velocity range between
$-35$ and $-90$ km\,s$^{-1}$ 
and for H\,{\sc i} column densities $>10^{19}$ cm$^{-2}$,
based on data of the H\,{\sc i} Leiden-Dwingeloo Survey.
Prominent IVC structures include the Intermediate-Velocity Arch (IV Arch; $l\sim 180, b\sim +70$),
the Low-Latitude Intermediate-Velocity Arch (LLIV Arch; $l\sim 160, b\sim +30$), 
the Intermediate-Velocity Spur (IV Spur; $l\sim 250, b\sim +60$),
and the Pegasus-Pisces Arch (PP Arch; $l\sim 110, b\sim -50$). From these
data it follows that IVC H\,{\sc i} gas with
$N$(H\,{\sc i}$)>10^{19}$ cm$^{-2}$ fills $\sim 40$ percent of the sky.\\
\indent A more detailed review of the distribution of HVCs and IVCs in the Milky Way halo is 
provided by Wakker (2004).

\begin{figure}[t!]
\epsfxsize=12.5cm
\epsfbox{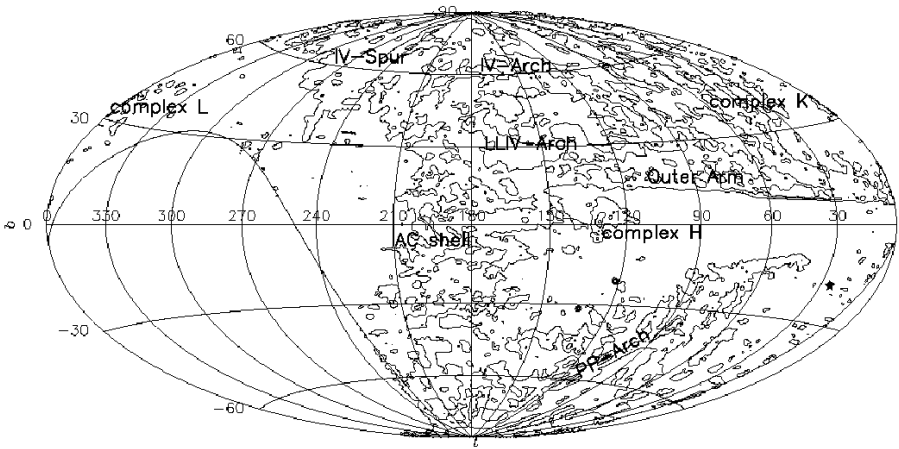}
\caption{
Aitoff projection all-sky map of the Galactic IVCs, in galactic
coordinates, for H\,{\sc i} column densities $>10^{19}$ cm$^{-2}$
and velocities between
$-35$ and $-90$ km\,s$^{-1}$,
based on the 21cm data from the Leiden-Dwingeloo Survey (LDS; Hartmann \& Burton 1997).
Individual IVC complexes are indicated. Figure taken from Wakker (2004).
[PLEASE CONTACT THE AUTHOR FOR A HIGH-QUALITY VERSION OF THIS FIGURE.]
}
\end{figure}

\subsection{Distances to IVCs and HVCs}

Measuring the distance to IVCs and HVCs is of great importance
to understand the nature and origin of these clouds, as the distance
information together with the sky distribution (see previous section)
provides a three-dimensional view of these clouds
around the Galaxy.
Unfortunately, very little information about IVC and HVC distances
is available. The reason for this is that reliable distance
measurements of IVCs and HVCs are very difficult and achievable only 
for a small number of these clouds. 

The most reliable (and only direct) method to measure IVC/HVC distances 
is the absorption-line method, in which one uses (mostly optical) high-resolution
spectra of stars with known distances in the direction of a halo cloud.
If a halo star is located behind an IVC or HVC that has a sufficiently
large gas-column density, one expects to see IVC/HVC absorption
(e.g., from the optical Ca\,{\sc ii} lines) 
in the stellar spectrum.
If the IVC/HVC is behind the star, no such absorption can occur.
One thus can bracket the distance of the IVCs and HVCs within a 
range defined by the distances of the available background stars. 
This method has been used for a few IVCs and HVCs.
IVCs with known distances are 
the IV Arch and the LLIV Arch (both have $d=0.5-2.0$ kpc); for some
other IVCs, upper distance limits of a few kpc have been derived
(see Wakker 2001 and references therein). The few HVCs for which
useful distance information is available are comlex A 
($d=4-10$ kpc; van Woerden et al.\,1999; Wakker 2001), 
complexes C and H ($d>5$ kpc; Wakker 2001), and complexes
WE ($d<13$ kpc; Sembach et al.\,1991) and WB ($d<16$ kpc; Thom et al.\,2005).
These values suggest 
that most IVCs represent objects that are relatively nearby
with typical distances of $<2$ kpc,
while many of the HVCs are more distant halo clouds with typical
distances $>5$ kpc. However, more of these distance measurements are required
to better understand the distribution of intermediate- and high-velocity gas 
surrounding the Milky Way. Recently, a very large number of 
blue-horizontal branch stars in the halo at high galactic latitudes
and in a large distance range ($d=0-90$ kpc) have been 
identified in various observational surveys 
(e.g., Brown et al.\,2004; Sirko et al.\,2004; Christlieb et al.\,2004).
These stars can be used as background sources
for Ca\,{\sc ii} absorption line measurements in the direction
of IVCs and HVCs, so that important new IVC/HVC distance information
is expected to become available during the next few years.

An indirect method to constrain distances of IVCs and HVCs is to
measure the intensity of H$\alpha$ emission from these clouds.
As recent observations demonstrate, several IVCs and HVCs shine bright
in H$\alpha$ emission (e.g., Weiner et al.\,2001; Putman et al.\,2003), 
implying that the neutral clouds are surrounded by 
ionized gaseous envelopes. If the Milky Way's ionizing radiation field
is responsible for the existence of these ionized gas layers, the
measured H$\alpha$ intensity in IVCs and HVCs serves as a measure for the distance
to the ionization source (i.e., the Galactic stellar disk). 
Weiner et al.\,and Putman et al.\,have
measured H$\alpha$ emission in a number of IVCs
and HVCs and conclude that many of these clouds are within
a radius of $\sim 40$ kpc from the Galaxy.
This therefore favors a
scenario in which most of the large IVCs and HVCs are located 
within the Milky Way halo but are not Local Group objects at
megaparsec distances.
A major problem with the H$\alpha$ method is, however, that next to 
radiation from the Milky Way disk collisional processes also 
may contribute to the ionization of the gaseous envelopes of these clouds.
This can be seen in the case
of the Magellanic Stream, which most likely represents gaseous
material torn out of the Magellanic Clouds during their tidal
interaction with the Milky Way. The Magellanic Stream is expected to be 
at least 50 kpc away and thus is too distant to be substantially
ionized by the Galaxy's radiation field. Yet, the Stream shows
H$\alpha$ emission at a level of up to $\sim 400$ mR (Putman et al.\,2003), 
suggesting that it has an ionized envelope that is produced by
collisional processes as the Stream is ramming into the hot, extended
Corona that surrounds the Milky Way (Sembach et al.\,2003).
Due to the unknown contribution of collisional processes 
to the ionization fraction in the envelopes of IVCs and HVCs,
distance estimates from H$\alpha$ fluxes are affected
by large systematic uncertainties.

An indirect method to obtain information about the distances
of the mysterious CHVCs
in the Milky Way halo is to consider the 21cm brightness temperature
of extraplanar and circumgalactic H\,{\sc i}
structures in M31, the other large spiral galaxy in the Local Group
that is $\sim 780$ kpc away (see also Section 4.1). 
Under the assumption that our sister
galaxy also comprises a population of H\,{\sc i} CHVCs with similar 
properties (i.e., radial distribution and emission characteristics mimic
those of the Milky Way CHVCs), the comparison of measured 21cm brightness
temperatures of the Galactic CHVCs with limits derived for the
M31 CHVC population at 780 kpc distance provides an estimate for the
distance of the CHVCs surrounding the Milky Way.
Using H\,{\sc i} 21cm data from the Effelsberg 100m radio telescope,
Westmeier et al.\,(2005a) conclude from the measured limits for the
M31 CHVC population that the Milky Way's CHVCs cannot have distances
larger than $\sim 60$ kpc. These measurements (just like the
H$\alpha$ observations discussed above) therefore suggest that the
Galactic H\,{\sc i} HVCs (including the CHVCs) are 
relatively nearby and thus represent a
{\it circumgalactic} rather than an {\it intergalactic/Local Group} 
phenomenon.

\begin{figure}[t!]
\epsfxsize=11.0cm
\epsfbox{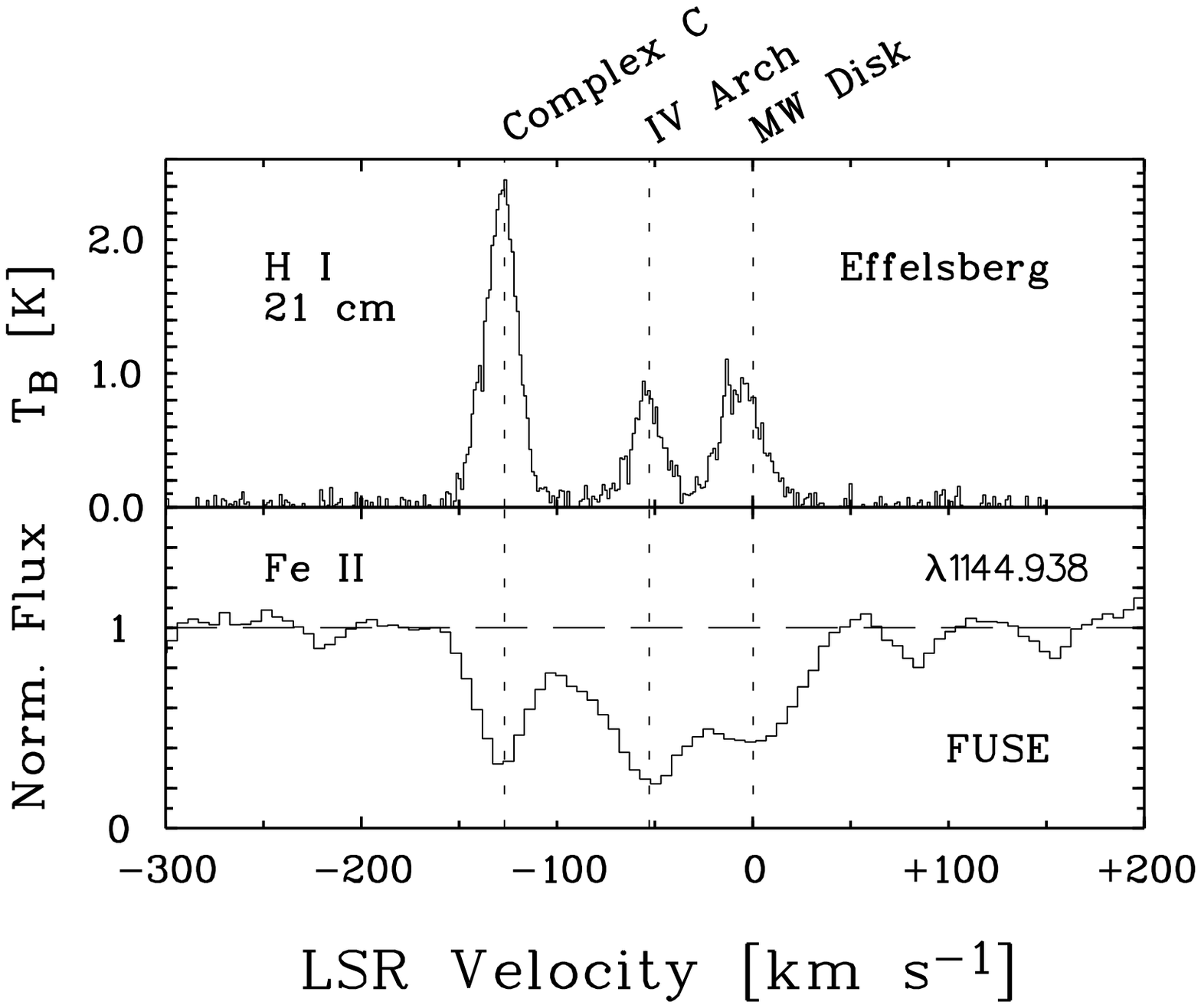}
\caption{
Velocity profiles of the H\,{\sc i} 21cm emission and 
Fe\,{\sc ii} $\lambda 1144.9$ FUV absorption 
in the direction of the quasar PG\,1259+593
($l=120.6$, $b=+58.1$).
Next to the local emission/absorption from Galactic disk gas
near zero velocities, IVC and HVC emission/absorption
is seen at $-55$ km\,s$^{-1}$ and $-130$ km\,s$^{-1}$,
indicating neutral halo gas associated with the IV Arch and HVC
complex C. Figure adapted from Richter et al.\,(2001b).
}
\end{figure}

\subsection{Metal Abundances and Origin of IVCs and HVCs}

Apart from the distance, the most valuable
information about the origin of IVCs and HVCs comes from 
studies of the chemical composition of these clouds.
Ultraviolet absorption-line spectroscopy
is the most sensitive and accurate method to measure
metal abundances and physical properties
in IVCs and HVCs, where typical gas
densities are significantly lower than in Galactic disk clouds.
The FUSE satellite has sufficient
sensitivity and spectral resolution to investigate absorption at
$\lambda \le 1187$ \AA\ in the Galactic halo and beyond along
a large number of sight lines. At longer wavelengths ($\lambda >
1150$ \AA), the STIS instrument installed on the 
Hubble Space Telescope (HST) provides additional information
about a number of atomic species and the Ly$\alpha$ absorption
line of neutral hydrogen near $1215.7$ \AA. Combining data from
these instruments thus provides a particularly powerful tool to
investigate metal abundances in intermediate- and high-velocity clouds.

During the last decade, a large number of measurements with FUSE, STIS,
and other instruments have provided 
important new results on the chemical composition
of the Galactic IVCs and HVCs (e.g., Richter et al.\,1999; 
Wakker et al.\,1999; Richter et al.\,2001a;
Richter et al.\,2001b; Collins et al.\,2003; Tripp et al.\,2003; 
Sembach et al.\,2004a).
These studies reveal
{\it disparate} chemical compositions for several IVCs and HVCs in different
directions in the sky, indicating that many of these clouds cannot 
have a common origin. Overall metal abundances in IVCs and HVCs 
typically vary between $\sim 0.1$ and $\sim 1.0$ solar. In general,
the chemical make-up of most of the IVCs and some of the HVCs can be
explained successfully by the galactic fountain model
(Shapiro \& Field 1976; Houck \& Bregman 1990), in which gas is ejected
out of the Galactic disk by supernova explosions and is falling back 
onto the disk in the form of neutral gas clouds.
Gas that is participating in this circulatory
pattern should have a nearly solar
metallicity reflecting that of their place of origin (i.e., the Milky Way disk).
Prominent examples for such solar-metallicity galactic fountain clouds are
the Intermediate-Velocity Arch (IV Arch; Richter et al.\,2001b), the Low-
Latitude Intermediate Velocity Arch (LLIV Arch; Richter et al.\,2001a), 
and HVC complex M (Wakker 2001).
There are, however, a number of HVCs whose abundances are inconsistent with
the Galactic fountain model.  One such case is the Magellanic Stream, which has
abundances close to those of the Small Magellanic Cloud (SMC; $\sim 0.3$ solar).
The Stream is believed to be tidally stripped
out of the SMC system during the last encounter with
the Galaxy (e.g., Wannier \& Wrixon 1972; Lu et al.\,1998;
Sembach et al.\,2001). High-velocity cloud complex C has an even lower
abundance ($\sim 0.1-0.3$ solar; Wakker et al.\,1999; Richter et al.\,2001b;
Tripp et al.\,2003) that is inconsistent with
gas originating in the disk of the Galaxy or in the Magellanic Clouds.
Thus, complex C might represent metal-poor material accreted from
the intergalactic medium. 
Possibly, mass shed by metal-poor halo red
giants contributes to the infall of low-metallicity gas, too
(de Boer 2004).
Accretion of substantial quantities of metal-poor gas
in the form of HVCs would have a significant influence on the chemical
evolution of the Milky Way. 

As an example for recent metal abundance measurements in Galactic IVCs and
HVCs we discuss the abundance pattern in IVC/HVC gas in the 
direction of the quasar PG\,1259+593 ($l=120.6$, $b=+58.1$). This 
sightline is particularly well suited to explore the metal content of IVC/HVC gas
with FUV absorption spectroscopy, as it passes {\it 
two different} Galactic halo clouds (one IVC and one HVC) that have different
radial velocities. The FUV spectrum of PG\,1259+593 therefore allows us
to directly compare the chemical abundances in two clouds using
the same spectral data. In addition, PG\,1259+593 lies in a direction in the 
sky where the local Galactic foreground gas in the disk has a very low
neutral gas column density. This is an advantage, since the many 
absorption lines from molecular hydrogen (H$_2$) in the disk 
gas are weak in this direction and thus H$_2$ line blending with IVC and HVC
components is much less severe than along other QSO sightlines.
Richter et al.\,(2001b) have analyzed  interstellar 
ultraviolet absorption lines in HVC complex C and the 
IV Arch in the direction of the quasar PG\,1259+593 using 
FUSE and STIS data. Fig.\,3 shows the velocity structure of 
interstellar gas in the direction of PG\,1259+593. In the
upper panel, the velocity profile of the H\,{\sc i} 21cm 
emission (based on observations with the Effelsberg 100m
radio telescope) is plotted against the LSR velocity. Next to
the local H\,{\sc i} 21cm emission from Galactic disk gas 
near zero velocities, IVC and HVC H\,{\sc i} emission 
is seen at $-55$ km\,s$^{-1}$ and $-130$ km\,s$^{-1}$, 
indicating halo gas related to the IV Arch and HVC
complex C. In the lower panel of
Fig.\,3 we show the absorption pattern in the Fe\,{\sc ii}
$\lambda 1144.938$ line, based on the FUSE observations of
PG\,1259+593. Fe\,{\sc ii} absorption by the IV Arch and
complex C are clearly visible and (despite the lower 
resolution) the velocity pattern perfectly
matches that of the H\,{\sc i} 21cm emission.
Other atomic species detected (in absorption) 
in the IV Arch and complex C include D\,{\sc i},
C\,{\sc ii}, N\,{\sc i}, N\,{\sc ii},
O\,{\sc i}, Al\,{\sc ii}, Si\,{\sc ii}, P\,{\sc ii}, 
S\,{\sc ii}, Ar\,{\sc i},
and Fe\,{\sc iii} (Richter et al.\,2001b; Sembach et al.\,2004a). 

\begin{figure}[t!]
\epsfxsize=12.0cm
\epsfbox{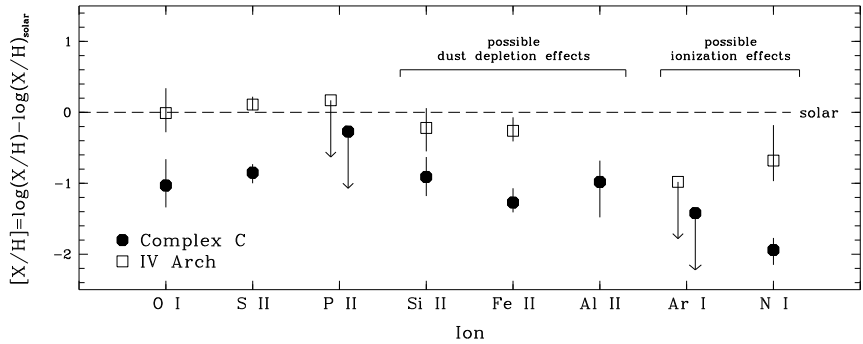}
\caption{
Normalized interstellar gas-phase abundances in complex C and the IV Arch.
Abundances in complex C are systematically lower than in the IV Arch,
suggesting a different enrichment history of both clouds. Figure
taken from Richter et al.\,(2001a).
}
\end{figure}

The O\,{\sc i}/H\,{\sc i}  ratio provides the best measure 
of the overall metallicity in the diffuse interstellar medium, 
because ionization effects do not alter
the ratio, and oxygen is at most only lightly depleted from 
the gas into dust grains.  For complex C, Richter et al.\,find an oxygen
abundance of 0.09 $^{+0.13}_{-0.05}$ solar,
consistent with the idea that complex C represents the 
infall of low-metallicity gas onto the Milky Way. In contrast,
the oxygen abundance in the IV Arch is 0.98 $^{+1.21}_{-0.46}$ solar,
which points to a Galactic origin. Similar abundance differences 
between the IV Arch and complex C are also found in other elements, but here, dust
depletion and ionization effects have to be taken into account for the
interpretation of the observed abundance ratios. 
The abundance pattern for various elements such as oxygen, sulfur,
phosphorus, silicon, iron, aluminum, argon, and nitrogen 
in the IV Arch and complex C is shown in Fig.\,4.
The abundance measurements toward PG\,1259+593 demonstrate that 
various different processes are responsible
for the phenomenon of intermediate- high-velocity neutral gas clouds in the
halo of the Milky Way - they cannot have a single origin.

An extensive summary of absorption line measurements in Galactic halo clouds 
is provided by Wakker (2001).

\subsection{Small-Scale Structure in IVCs and HVCs}

\begin{figure}[t!]
\epsfxsize=12.5cm
\epsfbox{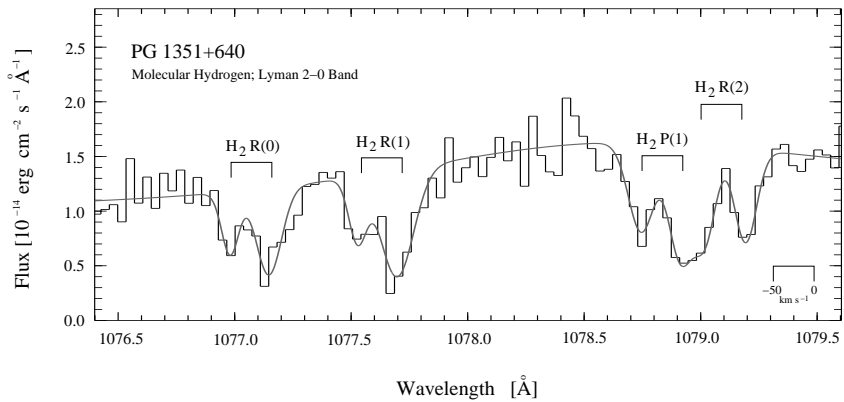}
\caption{
Excerpt of the FUSE spectrum of PG\,1351+640 in the wavelength range
between 1076.5 and 1079.5 \AA, sampling the H$_2$ $2-0$
Lyman band. The individual H$_2$ lines present in the data are
labeled above the spectrum. H$_2$ absorption from
the rotational levels $J=0,1$ and $2$ is seen in two components at $0$ and
$-50$ km\,s$^{-1}$, representing H$_2$ gas from the local Milky Way
and the IV Arch (core IV\,19), respectively. The solid gray line shows
a two-component Gaussian fit of the spectrum that
nicely reproduces the absorption pattern seen in the FUSE data.
Figure taken from Richter et al.\,(2003a).
}
\end{figure}

The gaseous halo of the Milky Way is an extreme multi-phase medium
with temperatures ranging from $50$ to several million
Kelvin. Although the large IVC and HVC complexes like
the IV Arch, complex C and the Magellanic Stream may span a depth of several 
kpc in the halo, small-scale structure in this gas is present
down to scales of several AU (e.g., Meyer \& Lauroesch 1999).
Studying this small-scale component in Galactic halo clouds 
not only provides an insight into the internal structure of 
IVCs and HVCs, but also yields important information about 
physical processes in the interstellar medium in general.   

Molecular hydrogen is an excellent diagnostic tool 
to investigate small-scale structure and 
physical conditions in IVCs and HVCs.
A large number of H$_2$ absorption lines from the Lyman and Werner band
is present in the FUV range
between $900$ and $1200$ \AA. 
The H$_2$ abundance in the diffuse ISM is
determined by a balance of the formation
of molecules on the surface of dust grains and the H$_2$ destruction
by the dissociating UV radiation (see Richter et al.\,2003a). 
For known photoabsorption and grain formation rates one
can estimate the hydrogen volume density
from the measured H\,{\sc i} and H$_2$ column densities. The
size of an H$_2$ absorbing structure, $D$, then can easily be calculated
from $N$(H\,{\sc i}) and $n_{\rm H}$.
Also the rotational excitation of the H$_2$ molecules can be
used to investigate physical conditions in the gas.
The lowest rotational energy states of H$_2$ (rotational
levels $J=0$ and $1$) are usually excited by collisions, so
that the column density ratio $N(1)/N(0)$ serves as a measure
for the kinetic temperature of the gas. In addition, the 
population of the higher rotational
states (determined by UV photon pumping) 
indicates the strength of the local UV radiation field.

A number of positive detections of H$_2$ absorption with FUSE and
ORFEUS has been reported for both IVCs and HVCs
(e.g., Richter et al.\,1999; Gringel et al.\,2000; Bluhm et al.\,2001;
Richter et al.\,2001c;
Sembach et al.\,2001; Richter et al.\,2003a).
Positive detections include the IV Arch, LLIV Arch, IV Spur, complex gp,
the Draco cloud, LMC-IVC, LMC-HVC, and the Magellanic Stream
(see also Richter \& de Boer 2004). 
In all cases, the observed column
densities are low (log $N$(H$_2)\leq 17$), implying
that the H$_2$ resides in a predominantly neutral gas phase.
As expected, H$_2$ absorption mostly occurs in
halo clouds that have a high metal and dust abundance,
thus preferably in IVCs rather than in HVCs (e.g., Richter et al.\,1999).
As an example for H$_2$ absorption in a Galactic IVC 
we show in Fig.\,5 the FUSE spectrum of the quasar
PG\,1351+640 in the range between $1076.5$ and $1079.5$ \AA,
where a number of H$_2$ lines from various rotational states
are present. Halo H$_2$ absorption at negative intermediate velocities
from gas in the IV Arch (core IV\,19)
is clearly visible in the various H$_2$ lines shown.

With FUSE, Richter et al.\,(2003a) have systematically studied
the properties of the H$_2$ gas in IVCs towards a large number
(56) of mostly extragalactic background sources.
The sample includes 61 IVC components with H\,{\sc i} column densities
$\geq 10^{19}$ cm$^{-2}$ and radial velocities
$25 \leq |v_{\rm LSR}| \leq 100$ km\,s$^{-1}$.
In FUSE spectra with good signal-to-noise
ratios (S/N$>8$ per resolution element) they find 14 clear
detections of H$_2$ in IVC gas with H$_2$ column densities
between $10^{14}$ and $10^{17}$ cm$^{-2}$.
In lower S/N data, H$_2$ absorption in IVC gas is tentatively
detected in an additional 17 cases. The molecular hydrogen fraction in
these clouds, $f=2N$(H$_2)/[N$(H\,{\sc i})$+2N($H$_2)]$, varies
between $10^{-6}$ and $10^{-3}$. This suggests that the H$_2$
lives in a relatively dense, mostly neutral gas phase that
probably is linked to the cold neutral medium (CNM) in these clouds.
Under the assumption of a H$_2$ formation-dissociation equilibrium one
can determine the
hydrogen volume density and the thickness of the absorbing structure.
The H$_2$ photoabsorption rate in the halo
depends on the mean ultraviolet radiation field at a height
$z$ above the Galactic plane. The models of Wolfire et al.\,(1995)
predict that that the radiation field and the 
H$_2$ photoabsorption rate at $\sim 1$ kpc above
the disk is approximately 50 percent of that within the disk.
If one assumes that the H$_2$ grain formation rate in
IVCs is roughly similar to that within the disk,
the H$_2$ and H\,{\sc i} column densities measured for the
Richter et al.\,IVC
sample imply mean hydrogen volume densities of $n_{\rm H}
\approx 30$ cm$^{-3}$ and linear diameters of the H$_2$
absorbing structures of $D\approx 0.1$ pc.
Moreover, if
one considers the rotational excitation of the halo H$_2$ gas
that can be measured for some of the IVC sightlines,
one finds for the kinetic temperature of this gas a conservative upper limit
of $T_{\rm kin} \leq 300$ K. Given the relatively high detection
rate of H$_2$ in these clouds, the measurements indicate that
the CNM phase in IVCs is ubiquitous and therefore 
represents a gas phase that is characteristic
for the denser, neutral regions in the halo. Most likely,
the CNM filaments are embedded in a more tenuous gas phase
that corresponds to the warm neutral medium (WNM).

The detection of H$_2$ absorption in a low-column density
IVC towards the LMC star Sk $-$68\,80 (Richter et al.\,2003b)
demonstrates that even smaller filaments at AU scales exist
in the halo. The observed gas clumps must have very high 
hydrogen volume densities (almost $10^3$ cm$^{-3}$) 
and relatively low gas temperatures ($T<50$ K). They 
probably are related to the so-called tiny-scale atomic 
structures (TSAS), small-scale structures that also have 
been found in the disk of the Milky Way (Heiles et al.\,1997;
Heithausen 2004). 
Many aspects
that concern the physical nature of these tiny filaments
(e.g., formation processes, thermal pressures,
dust content, etc.) are not well understood, and more data
are required to explore these intriguing objects in more detail.

\section{The Galactic Corona}

\subsection{Overview}

That the disk of our Milky Way is surrounded
by an envelope of hot gas was first proposed by Spitzer
(1956). Spitzer considered the presence of such a "Galactic Corona"
(according to the solar corona)
as necessary to explain spectroscopic observations that have been made
some years earlier by M\"unch (1952) and M\"unch \& Zirin (1961). They
had found interstellar Ca\,{\sc ii} absorption from neutral
gas clouds spread over a large velocity
range ($\sim 50$ km\,s$^{-1}$) towards O and B stars at high galactic
latitudes and large distances ($> 500$ pc) from the Galactic plane.
This was the first detection of the intermediate- and high-velocity clouds
in absorption.
Due to the low neutral gas density so far above the Galactic plane,
Spitzer argued that these clouds should not exist for long,
but disperse at time scales of $\sim 10^7$ years, unless they would
be embedded in a thin but hot, highly ionized gaseous medium that would provide
the necessary thermal pressure to confine these clouds. 
While the presence of the neutral high-velocity clouds
was confirmed shortly after by H\,{\sc i} 21cm radio
observations (Muller et al.\,1963; see Section 2.2),
it took about 20 years to find compelling observational
evidence for the hot coronal gas in which the IVCs and HVCs are embedded. 

In the mid-seventies,
observations with the Copernicus satellite (e.g., Jenkins
et al.\,1974) discovered the widespread presence of
five-times ionized oxygen (O\,{\sc vi}) within the disk
of the Milky Way. In addition, diffuse X-ray emission was found
toward high latitudes in rocket experiments (Williamson et al.\,1974).
To account for this newly detected ubiquitous hot interstellar
gas phase, and also for the high-velocity clouds that are falling towards
the Galactic disk, Shapiro \& Field (1976) developed the model
of the galactic fountain, in which the hot gas is produced
by supernova explosions in the disk of the Milky Way. This gas
forms cavities, which expand (due to overpressure) and
eventually break out of the disk. Hot gas then would stream into
the Milky Way halo several kiloparsecs above the disk, forming
a hot, gaseous corona of the Milky Way, as proposed by Spitzer
twenty years earlier. Part of this gas would be able to cool,
forming some of the intermediate- and high-velocity clouds. 
In the same way as for the neutral halo clouds, 
absorption line spectroscopy in the
UV and FUV is well suited to study hot gas
in the circumgalactic environment of the Milky Way, 
as the UV and FUV range contains a number of lines from 
highly-ionized species,
such as C\,{\sc iv}, N\,{\sc v}, and O\,{\sc vi}. These lines
sample gas in the temperature range between
$1\times 10^5$ and $5\times 10^5$ K and thus provide important
information about the distribution and physical properties 
of hot gas in the halo of the Milky Way (see also de Boer 2004).

\subsection{Distribution of O\,{\sc vi} in the Halo}

The first absorption-line studies of the Galactic Corona were
based on observations of Si\,{\sc iv}, C\,{\sc iv}, and
N\,{\sc v} with the {\it International
Ultraviolet Explorer} (IUE) 
and the HST (Savage \& de\,Boer 1979, 1981; Sembach \& Savage 1992; Savage et al.\,1997).
The best possible way to study the hot halo component of the Milky Way
is absorption spectroscopy of O\,{\sc vi},
which has as an ionization potential as large as $\sim 114$ eV.
O\,{\sc vi} traces either collisionally ionized, hot gas at
temperatures near $3\times10^5$ K, or low density gas that
is exposed to a very intense UV radiation field, or a mixture
of both. O\,{\sc vi} does not trace the very hot gas phase
(with temperatures exceeding $10^6$ K) that likely
exists in the coronal gas and that is responsible for the soft X-ray emission.
Gas at temperatures around $3\times10^5$ K cools
rapidly and O\,{\sc vi} absorption therefore is expected
to trace the interface regions between the hot ($10^6$ K) and
the warm ($10^4$ K) gas phase, e.g., in cooling flows,
cooling bubbles, and mixing layers.

A rather simple model to describe the density distribution of the hot gas as a
function of the vertical distance from the Galactic plane ($z$ height) is
to assume an exponential stratification. With $n_0$ as the
mid-plane gas volume density and $h$ as the
scale-height the gas-density at a given $z$
can be expressed as $n(z) = n_0\,{\rm exp}({-|z|/h})$. 
High-ion absorption line measurements in sightlines through the
halo provide a direct estimate of the stratification of the hot
gas away from the Galactic plane. For a given sightline in
the direction ($l,b$), the column density for the ion X, $N$(X),
is measured up to a height $|z|$ above the disk,
at which the background source is located 
(these background sources
may be halo stars or extragalactic objects).
Comparing $N$(X)\,sin\,$|b|$ with
$|z|$ for a large number of sightlines, one can fit an
exponential distribution (as shown above)
to the data points and derive the scale height $h$(X).
Such a $N$(X)\,sin\,$|b|$ vs.\,$|z|$
analysis has been done by Savage et al.\,(1997)
for the species Si\,{\sc iv}, C\,{\sc iv},
and N\,{\sc v}. Using spectra from the {\it Goddard
High Resolution Spectrograph} (GHRS) on HST they derive
exponential scale heights of
$h$(Si\,{\sc iv})$=5.1\pm0.7$ kpc,
$h$(C\,{\sc iv})$=4.4\pm0.6$ kpc,
and $h$(N\,{\sc v})$=3.9\pm1.4$ kpc. In contrast, for the
sample of objects the scale height derived for 
neutral hydrogen is just
$h$(H\,{\sc i})$=0.30\pm0.03$ kpc,
showing that the neutral gas phase 
is (with the exception of the IVCs and HVCs)
concentrated in a thin
layer in the Galactic disk.
Another possibility to estimate the scale height of the hot
gas in the halo is to analyze individual absorption
profiles of high ions assuming that the gas is co-rotating
with the underlying disk. Savage et al.\,(1997) show that
for C\,{\sc iv} such an analysis results in a
scale height of $\sim 4.5$ kpc, thus very similar to the 
one derived with the $N$(X)\,sin\,$|b|$ vs.\,$|z|$ method.

Using ORFEUS data, Widmann et al.\,(1998) have presented the
first systematic study of O\,{\sc vi} absorption
in the halo. With the availability of a large
number of FUSE absorption spectra from extragalactic
background sources in 1999 our knowledge about the
$\sim 3\times 10^5$ K gas component in the halo
(as traced by O\,{\sc vi} absorption) has improved
substantially.
Wakker et al.\,(2003), Savage et al.\,(2003), and
Sembach et al.\,(2003) present a large survey of
O\,{\sc vi} absorption along 102 lines of sight
through the Milky Way halo. They find strong
O\,{\sc vi} absorption in a radial-velocity range from
approximately $-100$ to $+100$ km\,s$^{-1}$
with logarithmic O\,{\sc vi} column densities 
(in units cm$^{-2}$) ranging from
$13.85$ to $14.78$ (Savage et al.\,2003).
At these radial velocities, the  O\,{\sc vi} absorbing
gas should be located in the thick disk and/or halo
of the Milky Way.
The distribution of the O\,{\sc vi} absorbing gas in the 
thick disk and halo is not uniform, but appears
to be quite irregular and patchy. A simple model assuming
a symmetrical plane-parallel
patchy layer of O\,{\sc vi} absorbing material
provides a rough estimate for the exponential O\,{\sc vi}
scale height in the halo. Savage et al.\,(2003)
find $h$(O\,{\sc vi}$)\sim2.3$ kpc with an $\sim 0.25$ dex
excess of O\,{\sc vi} in the northern Galactic polar region.
The correlation of O\,{\sc vi} with other ISM tracers,
such as soft X-ray emission, H$\alpha$, and H\,{\sc i}
21cm, is rather poor (Savage et al.\,2003).
Mixing of warm and hot gas and radiative
cooling of outflowing hot gas from
supernova explosions in the disk could explain
the irregular distribution of O\,{\sc vi}
absorbing gas in the halo of the Milky Way.

\subsection{Highly-Ionized High-Velocity Clouds}

O\,{\sc vi} absorption towards extragalactic background
sources is observed not only at radial velocities
$\leq100$ km\,s$^{-1}$, but also at
higher velocities (Wakker et al.\,2003; Sembach et al.\,2003).
These detections imply that next to the Milky Way´s hot "atmosphere"
(i.e., the Galactic Corona) 
individual pockets of hot gas exist that move 
with high velocities through
in the circumgalactic 
environment of the Milky Way. 
Such high-velocity O\,{\sc vi} absorbers may contain 
a substantial fraction of the baryonic matter in the
Local Group in the form of ionized hydrogen (e.g., Cen \& Ostriker 1999).

From their FUSE survey of high-velocity O\,{\sc vi} absorption
Sembach et al.\,(2003) find that probably more than 60 percent
of the sky at high velocities is covered by ionized hydrogen
(associated with the O\,{\sc vi} absorbing gas) above
a column density level of log $N$(H$^+)=18$, assuming
a metallicity of the gas of $0.2$ solar.
Some of the high-velocity O\,{\sc vi} detected with FUSE
appears to be associated with known high-velocity H\,{\sc i} 21cm
structures (e.g., the high-velocity clouds complex A,
complex C, the Magellanic Stream, and the Outer Arm).
Other high-velocity O\,{\sc vi} features, however, have no counterparts
in H\,{\sc i} 21cm emission. The high radial velocities for most of these
O\,{\sc vi} absorbers are incompatible with those expected
for the hot coronal gas (even if the coronal gas motion is
decoupled from the underlying rotating disk). A transformation
from the Local Standard of Rest to the Galactic Standard of
Rest and the Local Group Standard of Rest velocity reference frames
reduces the dispersion  around the mean of the high-velocity
O\,{\sc vi} centroids (Sembach et al.\,2003; Nicastro
et al.\,2003). This can be
interpreted as evidence that {\it some} of the O\,{\sc vi} high-velocity
absorbers are intergalactic clouds in the Local Group
rather than clouds directly associated with the Milky Way.
However, it is extremely difficult to discriminate between a Local
Group explanation and a distant Galactic explanation for these
absorbers.
The presence of intergalactic O\,{\sc vi} absorbing gas
in the Local Group is in line with
theoretical models that predict that there should be a large
reservoir of hot gas left over from the formation
of the Local Group (see, e.g., Cen \& Ostriker 1999).
However, further FUV absorption line measurements and
additional X-ray observations will be required to test
this interesting idea.

It is unlikely that the high-velocity O\,{\sc vi} is
produced by photoionization. Probably, the gas
is collisionaly ionized at temperatures of several
$10^5$ K. The O\,{\sc vi} then may be produced
in the turbulent interface regions between very hot
($T>10^6$ K) gas in an extended Galactic Corona
and the cooler gas clouds that are moving through
this hot medium (see Sembach et al.\,2003).
Evidence for the existence of such
interfaces also comes from the comparison of absorption
lines from neutral species like O\,{\sc i} with
absorption from highly-ionized species
like O\,{\sc vi} (Fox et al.\,2004).

\section{Circumgalactic Gas in other Galaxies}

\subsection{Extraplanar H\,{\sc i} Gas in Galaxies}

\begin{figure}[t!]
\epsfxsize=12.5cm
\epsfbox{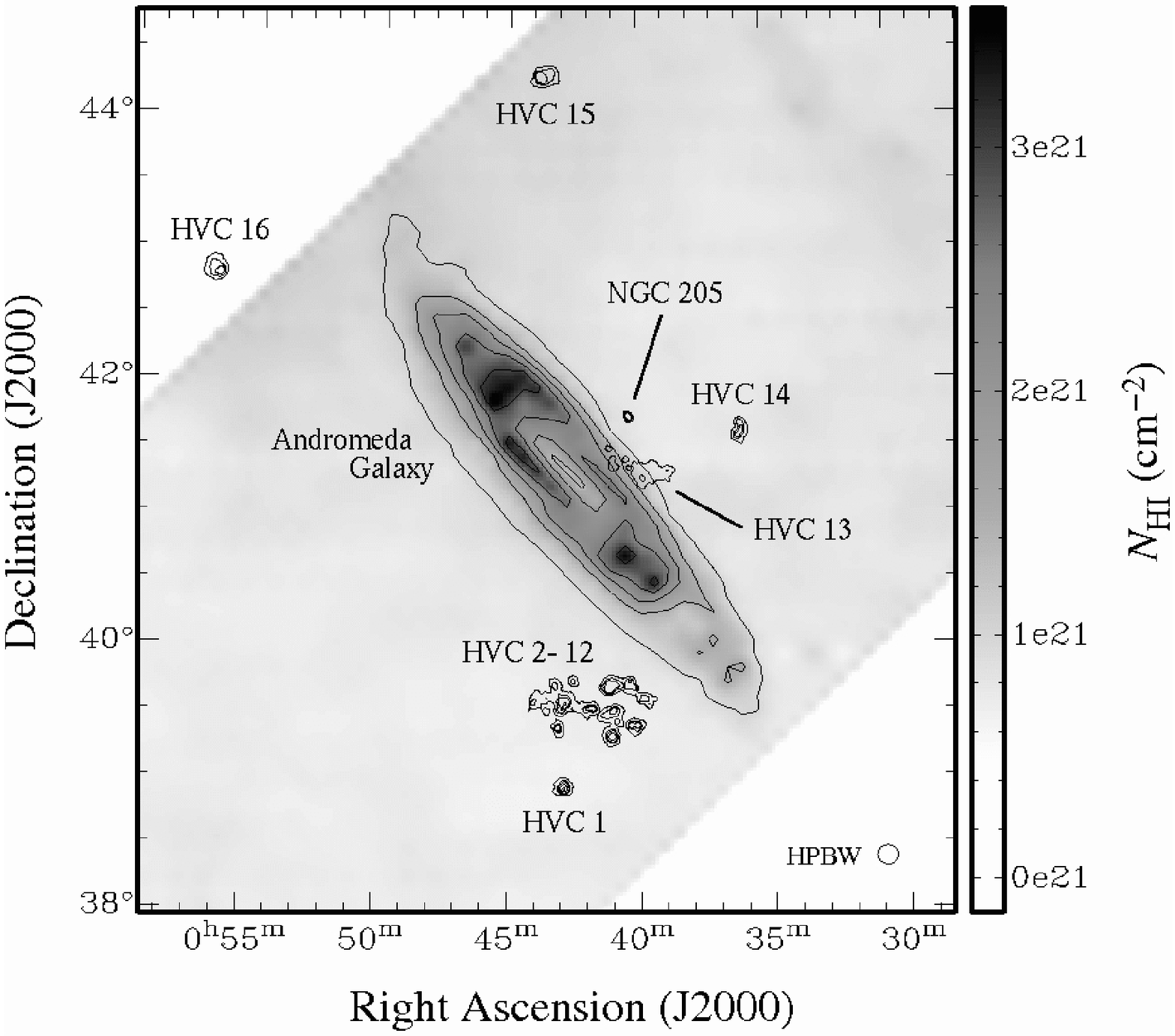}
\caption{
Distribution of H\,{\sc i} gas clouds in the halo of 
M31 (figure adapted from Westmeier et al.\,2005a, 
2005b). Several gas complexes are evident
at projected distances of several kpc above the disk of M31,
demonstrating that the second large spiral galaxy in the 
Local Group also possesses a population of high-velocity
clouds. HVCs and galaxies in the field are
labeled accordingly.
}
\end{figure}

The measurements of the Galactic IVCs and HVCs suggest that
circumgalactic neutral gas clouds play a fundamental role in the evolution
of the Milky Way. One thus would expect that also other galaxies
consist of such circumgalactic gaseous components, and that HVCs represent
a common phenomenon for spiral galaxies in the local Universe.
Finding HVCs in the halos of other galaxies is challenging, however,
due to the ambitious observational requirements, 
such as good mass sensitivity, 
a large field-of-view, and high spatial resolution. 
In recent years, great progress has been made to identify and
measure extraplanar neutral gas structures in other 
nearby galaxies. Positive detections of discrete extraplanar H\,{\sc i} clouds 
have been reported for the galaxies 
M31, M51, M81, M101, NGC\,891, and many others
(see Oosterloo 2004 for a review and references therein). 
Extensive H\,{\sc i} halos are found in several other cases. 
These structures extend
more than 10kpc away from the H\,{\sc i} disk of the galaxies and appear to
rotate slower than the underlying gaseous disk. Outflows as well
as tidal interactions with surrounding satellite galaxies both are believed to
contribute to the neutral gas flow in the halos of these galaxies.

A good example for HVCs in external galaxies is Andromeda (M31),
next to the Milky Way the other large spiral galaxy in the 
Local Group. Using the Green Bank Telescope, Thilker et al.\,(2004) 
have found around twenty H\,{\sc i} clouds in 21cm emission
in the halo of M31. They are located within $\sim 200$ km\,s$^{-1}$
of the systemic velocity of M31, thus in range similar to
what is found for the Milky Way HVCs. Follow-up observations
by Westmeier et al.\,(2005b) with the Westerbork Synthesis Radio
Telescope have confirmed the presence of the H\,{\sc i} HVCs around
M31. The derived H\,{\sc i} column densities ($\sim 10^{19}-10^{20}$
cm$^{-2}$) and H\,{\sc i} masses ($\sim 10^4-10^6\,M_{\odot}$) are
in line with values estimated for the Galactic HVC population.
Also the M31 HVCs most likely are confined by a hot gaseous corona.
Like the Milky Way, the Andromeda galaxy is surrounded
by a number of satellite galaxies. 
Morphological and kinematical properties of the M31 HVCs therefore suggest that 
some of these clouds have a tidal origin,
while other gaseous structures around M31 may represent fountain gas 
or material that is infalling from intergalactic space
(see Westmeier et al.\,2005b). Fig.\,6 shows the population of 
HVCs around M31, based on the data by Westmeier et al.\,(2005a, 2005b).

\subsection{Diffuse Hot Gas around Galaxies}

H$\alpha$ observations of edge-on spiral galaxies show the presence of
vertically extended layers of diffuse ionized gas (DIG) above
the disk. These measurements 
suggest that galactic-fountain processes (see Section 3) trigger the
flow of gas from the disk into the halo and back, and that 
these fountain processes are common in spiral galaxies. Studies of DIG
layers in the halos of galaxies at low redshift further indicate that 
their vertical extent and their brightness in the H$\alpha$ emission 
depend on the level of the underlying star formation in the disk 
(e.g., Rand 1996; Rossa \& Dettmar 2003), implying that the bulk 
of the ionized gas is produced by supernova explosions.
Therefore, ionized gas layers in the halos of disk galaxies trace
the disk-halo flow of the metal-enriched interstellar medium, 
while extraplanar H\,{\sc i} structures seen in these systems 
predominantly represent the return flow of this gas (after substantial 
cooling) and the infall of material from outside.  
Spiral galaxies are also found to possess highly structured clouds 
of absorbing dust extending to several kpc distances from their mid-planes 
(Howk \& Savage 1997, 1999).
The most  impressive example is the multiphase halo of NGC\,891, 
which exhibits numerous high-$z$ dust structures. 
Many of these structures contain
$\sim 10^5 \,M_{\odot}$ of gas and may be the sites
of star formation (Howk \& Savage 2000).

In addition to these H$\alpha$ and dust absorption observations, recent 
X-ray measurements of nearby edge-on galaxies with
Chandra and XMM-Newton have shown that these systems
consist of gigantic gaseous envelopes of hot, highly-ionized gas
(e.g., Wang et al.\,2001, 2005; Strickland et al.\,2004). 
These coronae most likely resemble the extended hot halo of 
our own galaxy. Near the disks (i.e., at small vertical 
distances from the galactic plane), X-ray emission and 
H$\alpha$ emission are strongly correlated. This supports the idea
that both methods
trace different temperature and density regimes of the same 
extraplanar ionized gas component.   
Also the X-ray luminosity of the coronal gas
in disk galaxies is proportional to the star-formation rate
in these systems (and to the total stellar mass). However, the observed
X-ray luminosities can account only for a small fraction of the 
expected supernova mechanical energy input, resulting in a
"missing energy" problem.
In late-type galaxies that are rich in cool gas much
of this missing energy is possibly radiated in the UV band. This
is supported by recent FUV emission-line observations (Otte et
al.\,2003), which indicate that emission in the two O\,{\sc vi}
lines at $\lambda\lambda 1031.9,1037.6$ may play an important role
for the overall energy balance in the hot gas.
In collisional ionization equilibrium, O\,{\sc vi} 
traces hot gas at temperature near $300,000$ K, thus
at the peak of the cooling curve of (metal-enriched) interstellar gas. 
At these temperatures, O\,{\sc vi} is expected to be the dominant 
coolant in the gas, and thus
O\,{\sc vi}-inferred cooling most likely can account
for a considerable fraction of the initial supernova energy input. 
However, 
as observations in our own halo indicate (e.g., Fox et al.\,2004; Section 3.3),
O\,{\sc vi} emission may preferentially arise in the interface regions
between the million-degree coronal gas and the neutral IVC and HVC gas.
Additional observations will be required to learn more about the 
distribution and physics of the O\,{\sc vi} emission in the halos of galaxies. 
Establishing a reliable estimate of the radiative cooling rate 
and unveiling the physical connection between FUV and X-ray emission 
will be of great importance to understand the physical conditions 
in the hot extraplanar gas of galaxies.

\subsection{Circumgalactic Absorption-Line Systems}

\begin{figure}[t!]
\epsfxsize=12.5cm
\epsfbox{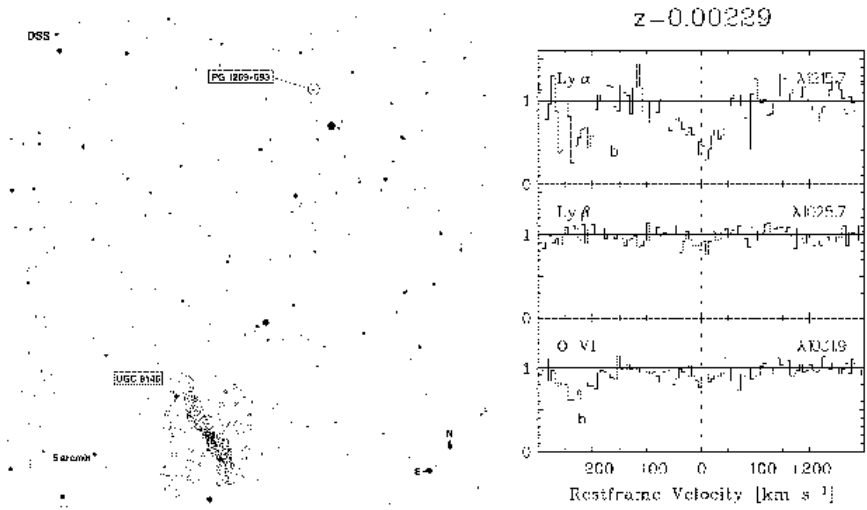}
\caption{
{\it Left panel:} 
sky positions of the quasar PG\,1259+593 and the galaxy UGC\,08146 
($z=0.00224$) on an image from the {\it Digitized Sky Survey}. 
H\,{\sc i} contours for
UGC\,08146 from data of the Westerbork Synthesis Radio Telescope
(Rhee \& van Albada 1996) are overlaid. {\it Right panel: } H\,{\sc i}
and O\,{\sc vi} absorption
profiles of the $z=0.00229$ absorption system toward PG\,1259+593.
The absorption probably arises from gas in the halo or immediate
intergalactic environment of UGC\,08146. Figure adopted from
Richter et al.\,(2004).
[PLEASE CONTACT THE AUTHOR FOR A HIGH-QUALITY VERSION OF THIS FIGURE.]
}

\end{figure}

The analysis of strong (log $N$(H\,{\sc i}$)\geq 15$)
quasar absorption-line systems that have associated metal 
lines (metal-line systems) is another important diagnostic
tool to investigate the halos of galaxies and their intergalactic
neighborhood at low and high redshifts. The presence of 
intervening absorption-line systems in QSO spectra can
be understood in terms of intergalactic gas filaments 
that trace baryon density fluctuations
related to gravitational instability during the
large-scale structure formation in the Universe
(see next Section). In this context, absorption-line
systems that have neutral gas column densities in the
range log $N$(H\,{\sc i}$)=15-19$ represent prime candidates for 
gaseous structures that reside
in the circumgalactic environment of galaxies:
the neutral gas column densities of metal-line systems 
are clearly higher than in the photoionized, low-density 
intergalactic medium, but they are
lower than characteristic column densities in the 
gaseous disks of present-day galaxies.

To find out whether or not an individual metal-line absorber
in a QSO spectrum is related to the halo of a galaxy, one first has
to identify candidate galaxies in the field of
the background quasar. In a second step, one has to compare
the absorber redshift with the redshift of the 
candidate galaxy to identify true (i.e., physically connected)
galaxy/absorber pairs.
Using this technique, many galaxy/absorber pairs 
have been found, implying that galaxy halos
extend far into intergalactic space, leaving
their imprint in the spectra of nearby (in terms
of angular separation) quasars. 
A particularly good tracer of metal-enriched 
gas that is associated with galaxy halos at
low redshift is the
strong resonant Mg\,{\sc ii} $\lambda\lambda 2796.4,2803.5$ 
doublet in absorption 
(e.g., Bergeron \& Stasinska 1986; Churchill et al.\,1999;
Churchill et al.\,2005). As an $\alpha$-process element,
magnesium is ejected into the interstellar and 
intergalactic gas by supernova explosions and thus
should be abundant in the halos of star-forming galaxies at 
low redshift. In addition, Mg\,{\sc ii} arises in gas that
spans roughly five orders of magnitudes in total gas
column densities (log $N$(H\,{\sc i}$)\approx 15.5-20.5$)
and thus is a sensitive tracer for circumgalactic gas that
may have very different ionization fractions and physical conditions.
Several studies of Mg\,{\sc ii} absorbers
in the fields of (absorption-selected) low-redshift 
galaxies have unveiled the connection between the 
galaxy luminosities, colors, and the extent of the 
Mg\,{\sc ii} absorbing envelopes 
around galaxies (e.g., Bergeron \& Boiss\'e 1991;
Steidel et al.\,1994). In general, the luminosities of
Mg\,{\sc ii} selected galaxies show only little evolution
with redshift, while the sizes of the Mg\,{\sc ii} halos 
scale only weakly with the luminosities of the 
galaxies (see, e.g., Churchill et al.\,2005). 
These measurements have shown that the gas cross section 
for strong Mg\,{\sc ii} absorbers with equivalent 
widths (EWs) $>0.3$ \AA\, is $\sim 40$ kpc, whereas for 
weaker systems with EWs $\leq 0.3$ \AA\, the cross
section is $\sim 70$ kpc. These values characterize
the sizes of Mg\,{\sc ii} halos of galaxies (that may 
have quite different morphologies). Note that the decrease 
of the Mg\,{\sc ii} equivalent width with increasing distance
to the galaxy does not necessarily indicate a diminishing
total gas column density. It rather implies
that the ionization fraction of the gas is higher at 
larger distances, as shown by observations of other, highly-ionized
species. It is therefore important to also consider other ions 
(e.g., C\,{\sc iv} and/or O\,{\sc vi} together
with H\,{\sc i}) for absorption-line studies of galaxy halos 
and the circumgalactic gaseous environment of galaxies.

As an example, we show in Fig.\,7 H\,{\sc i} and O\,{\sc vi}
absorption at $z=0.00229$ in the spectrum of the quasar
PG\,1259+593 ($z_{\rm em}=0.478$), based on FUSE observations
(Richter et al.\,2004). This sightline lies close to the
dwarf spiral galaxy UGC\,08146, which has a redshift of
$z=0.00224$. In Fig.\,7, left panel, we show
the sky positions of PG\,1259+593 and UGC\,08146 on an
image taken from the {\it Digitized Sky Survey} (DSS).
We have overlaid H\,{\sc i} 21cm contours
from data of the Westerbork Synthesis Radio Telescope
(WSRT; from Rhee \& von Albada 1996). Assuming
$H_0=75$ km\,s$^{-1}$\,Mpc$^{-1}$,
the redshift of $z=0.00224$
($672$ km\,s$^{-1}$) corresponds to a
distance of $\sim 9.0\,h^{-1}_{75}$ Mpc. Five arcmin in
Fig.\,7 therefore are equivalent to $\sim 13$ kpc
at the distance of UGC\,08146. The angular separation
between UGC\,08146 and the line of sight towards
PG\,1259+593 is $\sim 21$ arcmin,
so that the projected distance is $\sim 55\,h^{-1}_{75}$ kpc. 
The H\,{\sc i} and O\,{\sc vi} absorption at $z=0.00229$ in 
the spectrum of the quasar PG\,1259+593 therefore most likely 
arises in the halo or immediate intergalactic gaseous
environment of UGC\,08146.

The absorption-line characteristics of absorption-line systems around
galaxies in the low-redshift Universe can be compared with absorption
line systems arising the halo of our own galaxy. While most 
of the absorption measurements in the halo of the Milky Way concentrate
on the diagnostics of H\,{\sc i}-selected halo clouds (i.e., IVCs and
HVCs), Richter et al.\,(2005) have investigated Ca\,{\sc ii} absorption
in optical spectra of quasars that are distributed randomly in the sky.
Although the optical Ca\,{\sc ii} lines are less sensitive than the strong 
Mg\,{\sc ii} transition in the UV, they can be used to trace neutral 
high-velocity gas structures in the Milky Way halo at total gas column
densities below the detection limit of the large H\,{\sc i} HVC
surveys (e.g., the LAB survey, Kalberla et al.\,2005).
The big advantage of using Ca\,{\sc ii} absorption 
is that large ground-based telescopes such as the Very Large
Telescope (VLT) can be used
for such studies. Therefore, a large number of high S/N, high resolution  
spectra through the halo are available to study Ca\,{\sc ii}
absorption in the Galactic halo at high accuracy.
In contrast, Mg\,{\sc ii} and C\,{\sc iv} studies of gas in
the Galactic halo are limited in number and accuracy due to
the restricted sensitivity and availability of space-based
UV spectrographs. From a preliminary analysis of VLT/UVES spectra, 
Richter et al.\,(2005) find that weak Ca\,{\sc ii}  
absorption at high-velocities is seen in more than $\sim 50$ percent of  
the QSO spectra, also in directions where 
no corresponding H\,{\sc i} 21cm emission is seen in the large 
HVC 21cm surveys (e.g., Kalberla et al.\,2005). This implies that - next
to the well-known high-column density HVCs - the halo is filled with low-column
density neutral gas clouds that possibly represent the local 
counterparts of weak circumgalactic Mg\,{\sc ii} systems at low redshift 
that have equivalent widths of $\leq 0.3$ \AA.

\section{The Local Intergalactic Medium}

\subsection{Overview}

Shortly after the first detection of quasars in the 
early 1960s, the occurrence of many narrow absorption
lines in a QSO spectrum was recognized 
for the first time (see, e.g., Bahcall 1966). It soon became 
clear that these absorption lines are related to 
intervening gaseous structures that fill the intergalactic
space. These structures represent the intergalactic medium (IGM).

By far most of the intervening absorption lines 
in QSO spectra are produced
by H\,{\sc i} absorption in the Ly\,$\alpha$ line
($\lambda_0=1215.67$ \AA), which is redshifted to
higher wavelengths by the factor $(1+z)$, where $z$
denotes the redshift of an individual intergalactic 
absorber. With the availability of high-resolution
$N$-body hydrodynamical models (e.g., Dav\a'e et al.\,2001)
the Ly\,$\alpha$ absorbers have been interpreted as
gaseous structures that arise from baryon density fluctuations
associated with gravitational instability during the
large-scale structure formation in the Universe.
While the few strong absorption-line systems with H\,{\sc i} column
densities $>10^{15}$ cm$^{-2}$ (and associated metal
absorption) are believed to trace condensed structures such as
extended ionized galaxy halos, protogalaxies,
and gaseous disks of galaxies, the vast majority of the 
intervening absorption lines in QSO spectra 
are weak ($N$(H\,{\sc i}$)<10^{15}$ cm$^{-2}$).
In fact, the H\,{\sc i} column-density distribution function
of QSO absorption-line systems at high redshift
is (in a first-order approximation) a power-law
that scales with $N$(H\,{\sc i}$)^{-1.5}$ (e.g.,
Petitjean et al.\,1993).
Most of the weak intervening H\,{\sc i} lines belong 
to the so-called "Ly\,$\alpha$ forest",
which predominantly traces extended, highly-ionized
intergalactic structures - the "true" intergalactic 
medium. This gas has very low volume densities
($n_{\rm H}<10^{-4}$ cm$^{-3}$, typically) and is photoionized 
by the ambient UV background radiation from quasars and active galactic
nuclei. As structure evolution in the Universe proceeds
from high to low redshift,
however, an increasing fraction of the IGM is expected 
to be shock-heated and collisionally ionized, as the 
medium is collapsing under the action of gravity
in the deeper potential wells of the condensing
large-scale structure.
Since the intergalactic medium is highly ionized (the neutral gas fraction
is typically far less than a percent) it is clear that
the observed weak neutral hydrogen absorption serves only as a sign
for large amounts of ionized gas that is 
distributed in a filamentary network all over intergalactic space.

Following the results from IGM observations and numerical
simulations, photoionized and collisionally ionized intergalactic gas
most likely makes up for most of the baryonic matter in the local
Universe (where $\Omega_b\approx 0.045$).
While the diffuse photoionized IGM that
gives rise to the Lyman $\alpha$
forest accounts for $\sim 30$ percent of the
baryons today (Penton, Stocke, \& Shull 2004),
the shock-heated warm-hot intergalactic medium 
at temperatures $T\sim 10^5-10^7$ K is expected to contribute
at a comparable level to the cosmological mass density of
the baryons at $z=0$ (Cen \& Ostriker 1999; Dav\a'e et al.\,2001).
Gas and stars in galaxies, groups of galaxies, and galaxy
clusters make up the rest of the baryonic mass (Fukugita 2003).

\subsection{The Photoionized Ly\,$\alpha$ Forest at Low Redshift}

Absorption-line studies of the IGM at high redshift ($z>1.5$) represent
an essential observational method to directly study the structure 
evolution in the early Universe. Since all the important UV lines
are redshifted into the optical band, absorption-line 
spectroscopy of the high-$z$ IGM  with $8-10$m class telescopes
and state-of-the art spectrographs provide excellent spectral
data, i.e., spectra with high S/N ratios at high spectral resolution.
In contrast, measurements of the IGM at low $z$ require space-based
UV satellites and thus are much more challenging than for
high $z$. This fact leads to the somewhat unsatisfying situation
that many aspects of the local IGM and the IGM evolution from high
to low redshifts are not well understood yet. 

As measurements show, the number density (per unit redshift) 
of Ly\,$\alpha$ absorbers 
at high $z$ ($1.5\leq z \leq 4.0$) evolves rapidly, 
$dN/dz \propto (1+z)^{\gamma}$, where $\gamma \approx 2.2$ 
(e.g., Kim et al.\,2001). With the availability 
of HST, it therefore was a surprise to find that at
low redshifts ($z<1.5$) the evolution appears to be nearly
flat, with $\gamma \approx 0.1-0.3$ (Weymann et al.\,1998).
Note that these early low-$z$ observations of the 
Ly\,$\alpha$ forest were based on data obtained with 
the {\it Faint Object Spectrograph} (FOS), which has relatively
low spectral resolution. 
More recent, high-resolution HST/GHRS measurements (Penton et al.\,2002)
have shown, however, that the slowing of the 
Ly\,$\alpha$ density distribution in higher-resolution data
is not as great as previously measured in lower-resolution data. 
The break to slower evolution probably occurs at $z\sim 1.0$ 
rather than at $z\sim 1.5$. 
From these high-resolution GHRS 
measurements it also follows that higher-column 
density Ly\,$\alpha$ forest systems tend to 
cluster more strongly with galaxies than low-column density 
systems (Penton et al.\,2002). 
An example for a combined
FUSE/STIS high-resolution low-$z$ IGM spectrum toward 
the quasar PG\,1259+593 is shown in Fig.\,8. 

The evolution of the Ly\,$\alpha$ forest
density from high to low redshift can be understood in terms of
the expansion of the Universe and the changing ionizing background
flux. For redshifts from 5 to 1 the Ly\,$\alpha$ forest density
is decreasing rapidly due to the expanding Universe and the only 
mildly decreasing ionizing flux. At $z<1$ the evolution is
slowed down substantially due to the strongly decreasing 
ionizing background, which must have dropped by about one
order of magnitude from $z=1$ to $z=0$ (Haardt \& Madau 1996).
The line statistics together with a photoionization model of the
Ly\,$\alpha$ forest implies that the warm, photoionized intergalactic
medium at $z\approx 0$ contains $\sim 30$ percent of the 
baryons in the local Universe (Penton et al.\,2004). For comparison,
the contribution from the Ly\,$\alpha$ forest to the baryon density
($\Omega_b$) at $z=2$ is $\sim 90$ percent.
In the course of the structure evolution in the Universe from 
high to low redshifts the baryon budget in the 
Ly\,$\alpha$ forest therefore has dropped substantially. This is partly
due to the formation and assembly of condensed galactic structures,
but also due to the appearance of yet 
another intergalactic gaseous component,
the warm-hot intergalactic medium (WHIM).

\begin{figure}[t!]
\epsfxsize=12.5cm
\epsfbox{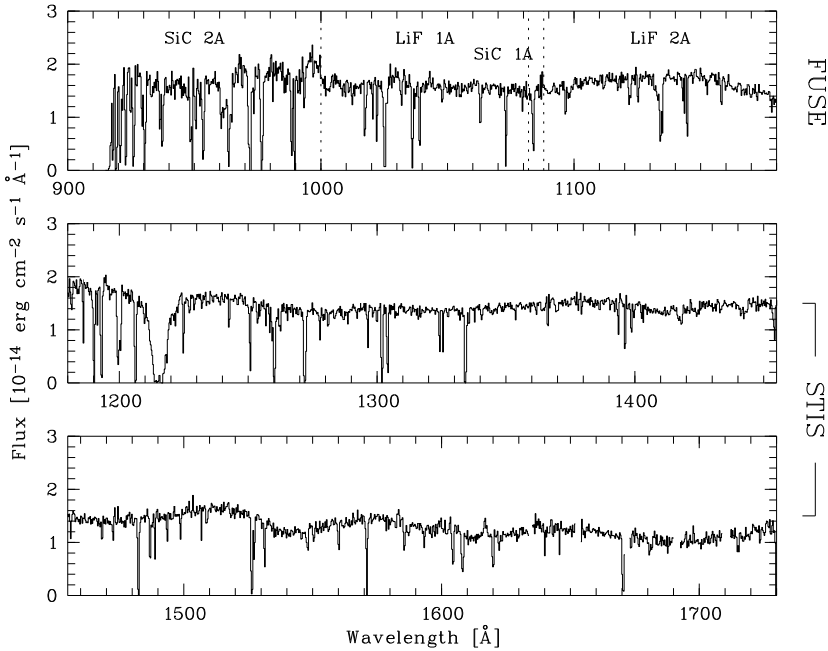}
\caption{
Combined FUSE and STIS spectrum of the low-redshift 
quasar PG\,1259+593 ($z_{\rm em}=0.478$) in the 
wavelength range between 900 and 1730 \AA. Next
to absorption from gas in the Milky Way, the spectrum
contains a number of low-$z$ intergalactic absorption
lines, as discussed in Richter et al.\,(2004).   
}
\end{figure}

\subsection{The Warm-Hot Intergalactic Medium}

As the temperature of the intergalactic medium
undergoes a significant change from
high to low redshifts, a large fraction of the baryonic matter
in the local Universe is expected to reside in the WHIM phase - a gas
phase that is particularly difficult to detect.
The WHIM is believed to emerge from intergalactic gas
that is shock-heated to high temperatures as the medium is
collapsing under the action of gravity (Valageas, Schaeffer, \& Silk 2002).
Directly observing this gas phase is challenging,
as the WHIM represents a low-density ($n_{\rm H}\sim10^{-6}-10^{-4}$ cm$^{-3}$),
high-temperature ($T\sim 10^5-10^7$ K) plasma that primarily is made of
protons, electrons, He$^+$, and He$^{++}$, together with traces of some
highly-ionized heavy elements. Diffuse emission from
this plasma is expected to have a very low surface brightness
and its detection awaits UV and X-ray
observatories more sensitive than currently available
(see, e.g., Fang et al.\,2005; Kawahara et al.\,2005).
A promising approach to
study the WHIM is the search for absorption features from the
WHIM in FUV and
in the X-ray regime. Five-times ionized
oxygen (O\,{\sc vi}) is the most important high ion
to trace the WHIM at temperatures of
$T\sim 3\times 10^5$ K
in the FUV regime. Oxygen is a
relatively abundant element and the two available O\,{\sc vi}
transitions (located at $1031.9$ and $1037.6$ \AA) have
large oscillator strengths.
A number of detections
of intervening WHIM O\,{\sc vi} absorbers at $z<0.5$
have been reported in the literature
(Tripp, Savage, \& Jenkins 2000; Oegerle et al.\,2000;
Chen \& Prochaska 2000; Savage et al.\,2002;
Richter et al.\,2004, Sembach et al.\,2004b; Savage et al.\,2005;
Danforth \& Shull 2005).
These measurements imply a number density of O\,{\sc vi}
absorbers per unit redshift of $dN_{\rm OVI}/dz \approx 17\pm 3$
for equivalent widths $W_{\lambda}\geq30$ m\AA\,(Danforth \& Shull 2005).
Assuming that 20 percent or less of the oxygen is present
in the form of O\,{\sc vi} ($f_{\rm O\,VI}\leq0.2$)
and further assuming a mean oxygen abundance of $0.1$ solar,
the measured number density of O\,{\sc vi} absorbers
corresponds to a cosmological mass density
of $\Omega_b$(O\,{\sc vi})$\geq0.0024$ $h_{75}\,^{-1}$.
Fig.\,9 shows an example of an intervening 
O\,{\sc vi} system at $z=0.31978$ in the direction
of PG\,1259+593.
For the interpretation of $\Omega_b$(O\,{\sc vi}) 
it has to be noted
that O\,{\sc vi} absorption traces 
collisionally ionized gas at
temperatures around $3 \times 10^5$ K (and also 
low-density, photoionized gas at lower temperatures), but not the
million-degree gas phase which probably contains
the majority of the baryons in the WHIM.
Very recently, Savage et al.\,(2005) have reported the
detection of Ne\,{\sc viii} in an absorption system
at $z\approx 0.2$ in the direction of the quasar HE\,0226$-$4110.
Ne\,{\sc viii} traces gas at $T\sim 7\times10^5$ K (in collisional
ionization equilibrium) and thus
is possibly suited to complement the O\,{\sc vi} measurements
of the WHIM in a higher temperature regime.
However, as the cosmic abundance of Ne\,{\sc viii} is relatively low,
Ne\,{\sc viii} is not expected to be a particularly sensitive tracer of the
WHIM at the S/N levels achievable with current UV
spectrographs.
This is supported by the non-detections of intervening Ne\,{\sc viii}
in other high S/N STIS data (Richter et al.\,2004).
Also X-ray absorption
measurements are very important for studying the WHIM 
(e.g., Fang et al.\,2002; Nicastro et al.\,2005), but they 
currently are limited in scope because of the small number of
available background sources and the relatively low spectral
resolution of current X-ray observatories (FWHM$\sim500$ to
$1000$ km\,s$^{-1}$). 

\begin{figure}[t!]
\epsfxsize=10.0cm
\epsfbox{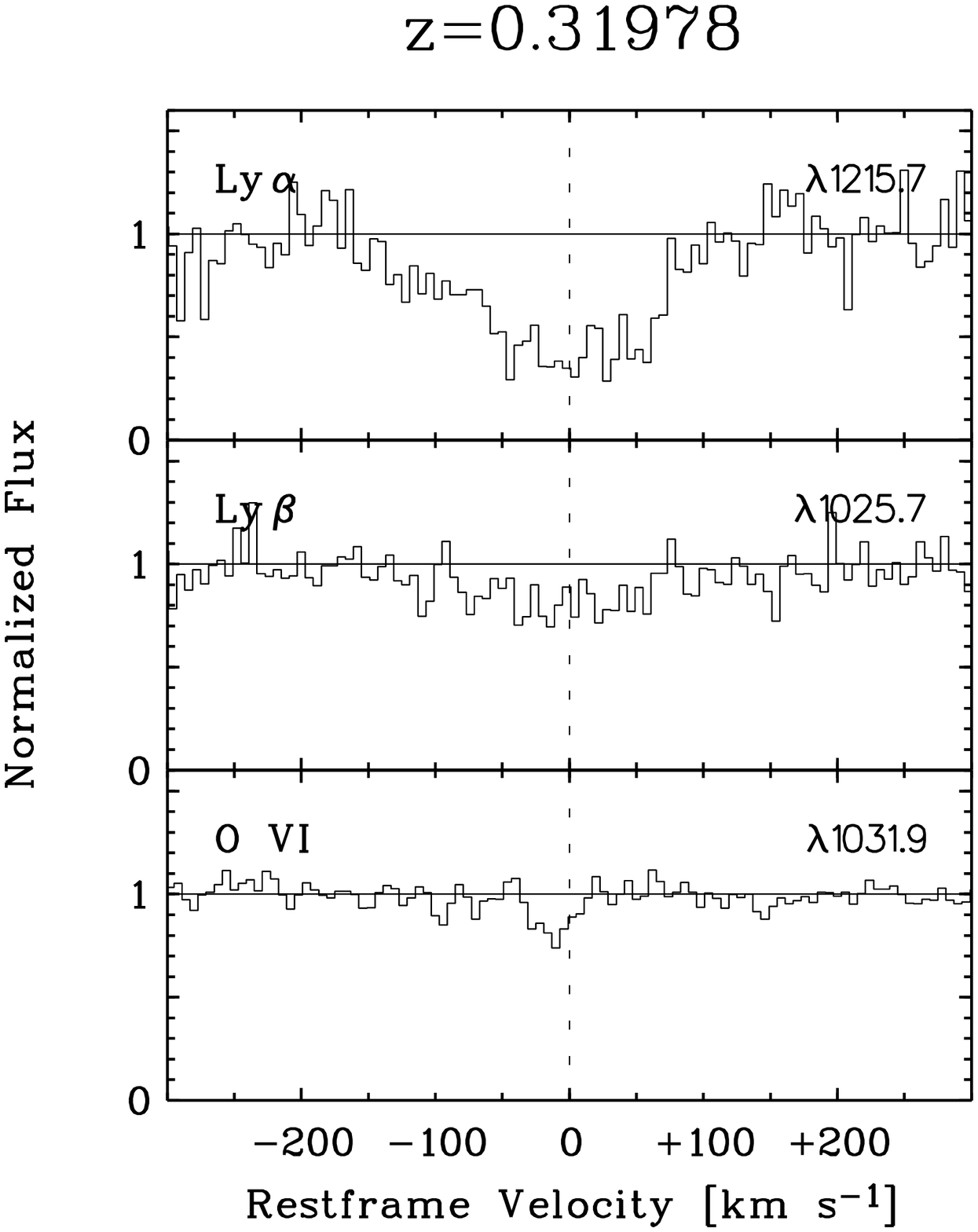}
\caption{
Normalized absorption profiles of H\,{\sc i} and O\,{\sc vi} 
in the O\,{\sc vi}/BLA system at $z=0.31978$ toward PG\,1259+593. 
Both the occurrence of O\,{\sc vi} absorption and the 
large width of the H\,{\sc i} absorption (probably caused
by thermal line broadening) are consistent with 
collisionally ionized gas at a temperature near $3\times 10^5$ K.
The total gas column density in this system is estimated to
be $\sim 6.4 \times 10^{19}$ cm$^{-2}$.
Figure and numbers taken from Richter et al.\,(2004).
}
\end{figure}

Next to high-ion absorption from oxygen and other metals,
recent observations with STIS (Richter et al.\,2004; Sembach
et al.\,2004b) have shown that WHIM filaments can be detected
in Ly\,$\alpha$ absorption of neutral hydrogen.
Although the vast majority of the hydrogen in the WHIM
is ionized (by collisional processes and UV radiation), a
tiny fraction ($f_{\rm H\,I}<10^{-5}$, typically) of neutral
hydrogen is expected to be present. Depending on the total
gas column density of a WHIM absorber and its
temperature, weak H\,{\sc i} Ly\,$\alpha$ absorption
at column densities $12.5\leq$ log $N$(H\,{\sc i})$\leq 14.0$
may arise from WHIM filaments and could be used to
trace the ionized hydrogen component.
The Ly\,$\alpha$ absorption from WHIM filaments is 
very broad due to thermal line
broadening, resulting in large Doppler parameters
of $b>40$ km\,s$^{-1}$. Such lines are generally difficult
to detect, as they are broad and shallow. High resolution,
high S/N FUV spectra of QSOs with smooth background continua
are required to successfully search for broad Ly\,$\alpha$
absorption in the low-redshift WHIM. STIS installed on the HST
is the only instrument that has provided such data,
but due to the instrumental limitations of space-based
observatories, the number of QSO spectra adequate for searching
for WHIM broad Ly\,$\alpha$ absorption 
(in the following abbreviated as "BLA") is very limited. An 
example for a BLA at $z=0.18047$ in the STIS spectrum of the quasar H\,1821+643
is shown in Fig.\,10.
So far, four sight lines observed with STIS
towards the quasars PG\,1259+593
($z_{\rm em}=0.478$), PG\,1116+215 ($z_{\rm em}=0.176$),
H\,1821+643 ($z_{\rm em}=0.297$), and PG\,0953+415 ($z_{\rm em}=0.239$)
have been carefully inspected for the presence of
BLAs, and a number of good candidates
have been identified 
(Richter et al.\,2004, 2006a; Sembach et al.\,2004).
These measurements imply a BLA
number density per unit redshift
of $dN_{\rm BLA}/dz \approx 22-53$ for Doppler parameters
$b\geq40$ km\,s$^{-1}$ and above a sensitivity limit of 
log ($N$(cm$^{-2})/b($km\,s$^{-1}))\geq 11.3$.
The large range for $dN_{\rm BLA}/dz$ partly is due to the uncertainty
about defining reliable selection criteria for
separating spurious cases from good broad Ly\,$\alpha$ candidates
(see discussions in Richter et al.\,2004, 2006a and Sembach et al.\,2004).
Transforming the number density $dN_{\rm BLA}/dz$ into a cosmological
baryonic mass density, one obtains $\Omega_b$(BLA)$\geq 0.0029\,h_{75}\,^{-1}$.
This limit is about 6 percent of the total baryonic mass density
in the Universe expected from the current cosmological models (see above), 
and is slightly above the limit derived for the intervening O\,{\sc vi} absorbers 
($\Omega_b$(O\,{\sc vi})$\geq 0.0024$ $h_{75}\,^{-1}$;
see above). The analysis of BLAs in cosmological simulations 
(Richter et al.\,2006b) supports the idea that
these systems represent a huge baryon reservoir in the local
Universe.

\begin{figure}[t!]
\epsfxsize=13.0cm
\epsfbox{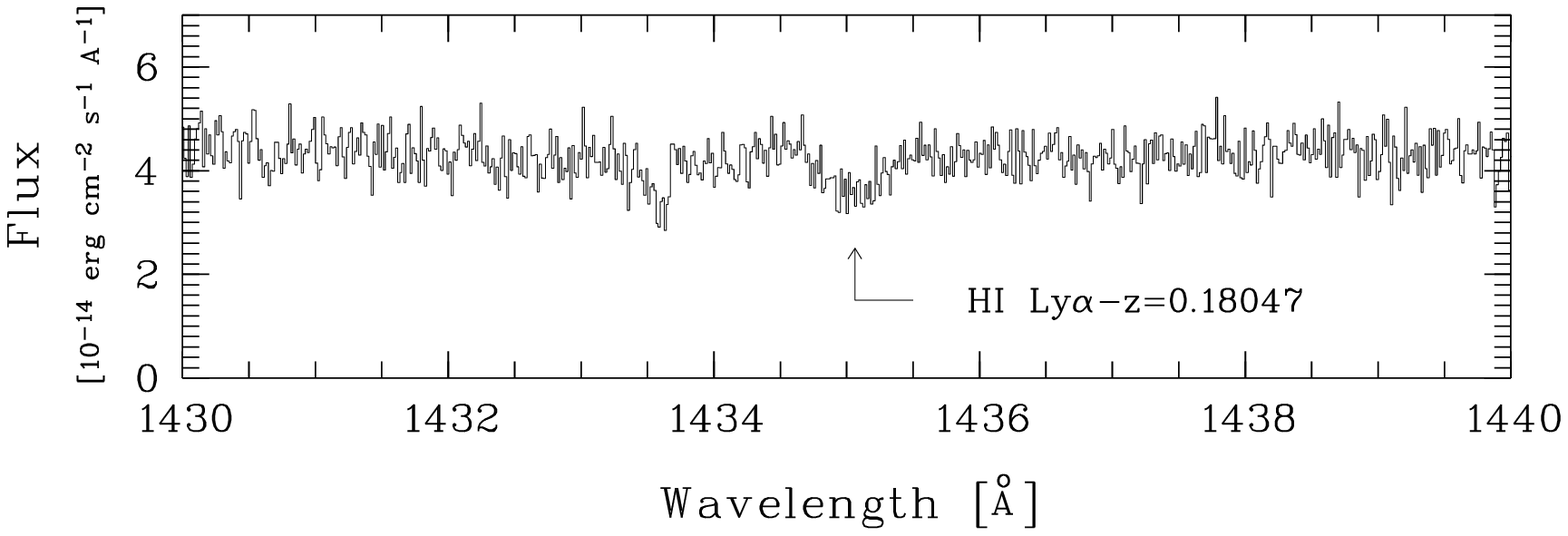}
\caption{
The well-detected broad Ly\,$\alpha$ absorber (BLA)
at $z=0.18047$ in the STIS spectrum of H\,1821+643 is shown.
The shape of this BLA differs significantly from the shape 
of the narrow lines from the Ly\,$\alpha$ forest, such 
as the $z=0.17924$ Ly\,$\alpha$
forest absorber near $1433.5$ \AA. Figure taken from 
Richter et al.\,(2005b).
}
\end{figure}

\subsection{Distribution of Metals}

As the many detections of metal lines in intervening
absorption-line systems show, the intergalactic medium contains
heavy elements. These elements must have been produced 
in stars within galaxies and then
transported through an efficient mechanism into the IGM.
The prime candidate for the metal enrichment of the IGM is outflows
from starbursting dwarf galaxies (e.g., Heckman et al.\,2001). Recent
studies using FUSE absorption-line data 
imply that the metallicity of the low-$z$ IGM is $\sim 0.1$ solar.
For instance, the number density of intervening 
O\,{\sc vi} systems observed toward low-redshift QSOs is in agreement
with the predicted distribution of O\,{\sc vi} absorbers 
in numerical simulations that assume an oxygen abundance of 0.1 solar
(Danforth \& Shull 2005). Also the measurements of intergalactic C\,{\sc iii} 
together with appropriate multi-phase ionization models suggest
a mean metallicity of the local IGM that is roughly ten percent solar
(Danforth et al.\,2005). 

While all these studies aim at 
providing an accurate {\it mean} metallicity of the local
IGM, it has to be kept in mind that metal abundances in 
individual intergalactic regions may
deviate substantially from this mean IGM metallicity. 
An example for an O\,{\sc vi} system with an oxygen abundance that
probably is much less than 0.1 solar is the absorber at $z=0.31978$
towards the quasar PG\,1259+593 (Richter et al.\,2004; see Fig.\,9). 
Due to its simplicity and its suggested high-temperature characteristics,
this system represents an interesting case for a more detailed
modeling of the physical conditions and the oxygen abundance.
The large H\,{\sc i} $b$ value
and the presence of O\,{\sc vi} already suggest
that this absorption system consists mainly
of hot gas at temperatures around $3\times10^5$ K.
Using the absorption width as a measure for the 
temperature of the gas, the collisional ionization
model implies hat the O abundance in this O\,{\sc vi} system is
only $\sim 4.3\times10^{-3}$ solar. 
An abundance as low as $\sim 4 \times10^{-3}$ solar, if correct,
has important implications since the 
estimated baryonic content of the O\,{\sc vi}
systems scales inversely with the assumed oxygen abundance.
Various groups (e.g., Savage et al.\,2002; Sembach et al.\,2004b;
Danforth \& Shull  2005) have estimated 
that the gas in low redshift O\,{\sc vi} systems
contributes with $\sim0.002$ to the cosmological closure density assuming that the
average oxygen abundance in O\,{\sc vi} systems is $0.1$ solar. This
contribution is comparable to that found in galaxies but $\sim20$ times
smaller than the total contribution for baryons estimated from the
Cosmic Background radiation or from big bang nucleosynthesis. However,
if the typical oxygen abundances in some of the O\,{\sc vi} systems are 10
to 25 times smaller, the estimate for the baryonic content of these
systems would increase by factors of 10 to 25. Finding low-metallicity
regions in the local IGM and studying their ionization properties 
therefore is of great importance to
provide reliable estimates of the baryon budget in WHIM
O\,{\sc vi} absorbers. 

\section{Conclusions and Outlook}

The results discussed in this article demonstrate 
that studies of the distribution and physical properties
of the gaseous circumgalactic and 
intergalactic environment of galaxies are important
to understand the formation and evolution of individual 
galaxies and the large-scale structure formation in
the Universe. The use of the absorption-line technique
in the FUV and optical in combination with emission 
measurements at X-ray, UV, optical, and radio wavelengths provides  
a particularly powerful to study the gaseous 
environment of galaxies and the intergalactic 
medium at low redshift. Concerning our own galaxy,
such combined absorption and
emission measurements have shown that 
the formation of the Milky Way is still on-going.
Other studies have demonstrated that the gaseous 
environment of galaxies and the intergalactic medium is 
(in general) governed by a complex interplay 
between gaseous infall and outflow and  
the hierarchical merging of (galactic) structures
in the local Universe. Despite the low densities that
characterize the gas in the outskirts of galaxies and
in the intergalactic medium, it has become clear that
(due to the large filling factor)
this gas contains a substantial 
fraction of the baryonic mass in the local Universe.

\subsection*{Acknowledgements}

First of all, I thank the Astronomische Gesellschaft for
awarding me the Ludwig Biermann Price - I am deeply honored.
Clearly, this price serves as an inspiration and motivation for
my future scientific work.   
I particularly like to thank my former supervisors 
Klaas de Boer and Blair Savage, who greatly
supported me over the last few years and from whom I
learned so much. I am grateful to my other past and recent 
collaborators and supporting colleagues Jacqueline
Bergeron, Hartmut Bluhm, Christian Br\"uns, Greg Bryan, Ralf-J\"urgen Dettmar, 
Taotao Fang, Andrea Ferrara, Andrew Fox, Gerhard Hensler, Michael Hilker,
Hiro Hirashita, Chris Howk, Peter Kalberla, Cedric Ledoux, Ole Marggraf,
Patrick Petitjean, Tom Richtler, Peter Schneider, Ken Sembach, Todd Tripp, Bart Wakker, 
Tobias Westmeier, and many others. It is a great pleasure to work 
together with so many bright and inspiring scientists.
Also, I would like to thank Klaas de Boer, Blair Savage, 
and Tobias Westmeier 
for a critical reading of the manuscript. Last but not least, 
I thank the {\it Deutsche Forschungsgemeinschaft} (DFG)
for financial support in the DFG Emmy-Noether program.

\subsection*{References}

{\small

\bref
Bahcall, J.N. 1966, ApJ 145, 684

\bref
Bergeron, J., Stasinska, G. 1986, A\&A 169, 1

\bref
Bergeron, J., \& Boiss\'e, P. 1991, A\&A 243, 344

\bref
Blitz, L., Spergel, D.N., Teuben, P.J., Hartmann, D., 
\& Burton, W.B. 1999, ApJ 514, 818

\bref
Bluhm, H., de\,Boer, K., Marggraf, O., \& Richter, P. 2001, A\&A 367, 299

\bref
Braun, R., \& Burton,, W.B. 1999, A\&A 341, 437

\bref
Braun, R., \& Burton,, W.B. 2000, A\&A 354, 853

\bref
Brown, W.R., Geller, M.J., Kenyon, S.J., et al. 2004, AJ 127, 1555

\bref
Cen, R., \& Ostriker, J. 1999, ApJ 514, 1

\bref
Chen, H.-W., \& Prochaska, J.X. 2000, ApJ 543, L9

\bref
Christlieb, N., Beers, T.C., Barklem, P., et al.\,2004, A\&A 428, 1027

\bref
Churchill, C.W., Rigby, J.R., Charlton, J.C., Vogt, S.S. 1999, ApJS 120, 51

\bref
Churchill, C.W., Steidel, C.C., Kacprzak, G. 2005, 
in Extra-planar gas, ed. R. Braun, ASP Conference Series,  Vol.\,331, 387

\bref
Collins, J.A., Shull, J.M., \& Giroux, M.L. 2003, ApJ 585, 336

\bref
Danforth, C.W., \& Shull, J.M. 2005, ApJ 624, 555

\bref
Danforth, C.W., Shull, J.M., Rosenberg, J.L., \& Stocke, J.T. 2005, astro-ph 0508656

\bref
Dav\'e, R., Cen, R., Ostriker, J., et al.\,2001, ApJ 552, 473

\bref
de Boer, K.S. 2004, A\&A 419, 527

\bref
de Boer, K.S. 2004, 
in High-velocity Clouds, ed. H. van Woerden et al.,
ASSL, Vol.\,312 (Kluwer), ISBN 1402025785, 227

\bref
Fang, T., Marshall, H.L., Lee, J.C., Davis, D.S., \& Canizares, C.R. 2002,
ApJ 572 L127

\bref
Fang, T., Croft, R.A.C., Sanders, W.T., et al. 2005, ApJ 623, 612

\bref
Fox, A.J., Savage, B.D., Wakker, B.P., et al. 2004, ApJ 602, 738

\bref
Fukugita, M. 2003, astro-ph 0312517

\bref
Gringel, W., Barnstedt, J., de\,Boer, K.S., et al. 2000, A\&A 358, L37

\bref
Haardt, F., \& Madau, P. 1996, ApJ 461, 20

\bref
Hartmann, D., \& Burton, W.B. 1997, Atlas of Galactic Neutral Hydrogen,
Cambridge University Press

\bref
Heckman, T.M., Sembach, K.R., Meurer, G.R., et al. 2001, ApJ 554, 1021

\bref
Heiles C. 1997, ApJ 481, 193

\bref
Heithausen, A. 2004, ApJ 606, L13

\bref
Howk, J.C., \& Savage, B.D. 1997, AJ 114, 2463

\bref
Howk, J.C., \& Savage, B.D. 1999, AJ 117, 2077

\bref
Howk, J.C., \& Savage, B.D. 2000, AJ 119, 644

\bref
Houck, J.C., \& Bregman, J.N. 1990, ApJ 352, 506

\bref
Hulsbosch, A.N.M., Wakker, B.P. 1988, A\&AS 75, 191

\bref
Ibata, R.A., Gilmore, G., \& Irwin, M.J. 1994, {\it Nature} 370, 194

\bref
Kalberla, P. M. W., Burton, W. B., Hartmann, D., et al.\,2005, A\&A 440, 775

\bref
Kawahara, H., Yoshikawa, K., Sasaki, S., et al. 2005, astro-ph 0504594

\bref
Kim, T.-S., Christiani, S., \& D'Odorico, S. 2001, A\&A 373, 575

\bref
Lu, L., Sargent, W.L.W., Savage, B.D., et al. 1998, AJ 115, 162

\bref
Meyer D.M. \& Lauroesch J.T. 1999, ApJ 520, L103

\bref
Morras, R., Bajaja, E., Arnal, E.M., \& P\"oppel, W.G.L. 2000, A\&AS 142, 25

\bref
Muller, C.A., Oort, J.H., \& Raimond, E. 1963, C.R. Acad. Sci. Paris 257, 1661

\bref
M\"unch, G. 1952, PASP 64, 312

\bref
M\"unch, G., \& Zirin, H. 1961, ApJ 133, 11

\bref
Murphy, E.M., Lockman, F.J., \& Savage, B.D. 1995, ApJ 447, 642

\bref
Nicastro, F., Zezas, A., Elvis, M., et al. 2003, {\it Nature} 421, 719

\bref
Nicastro, F., Mathur, S., Elvis, M., al. 2005, {\it Nature} 433, 495

\bref
Oegerle, W.R., Tripp, T.M., Sembach, K.R., et al.\,2000, ApJ 538, L23

\bref
Oort, J.H. 1970, A\&A 7, 381

\bref
Oosterloo, T. 2004, in High-velocity Clouds, ed. H. van Woerden et al.,
ASSL, Vol.\,312 (Kluwer), ISBN 1402025785, 125

\bref
Otte, B., Murphy, E.M., Howk, J.C., et al. 2003, ApJ 591, 821

\bref
Penton, S.V., Stocke, J.T., \& Shull, J.M. 2004, ApJS 152, 29

\bref
Penton, S.V., Stocke, J.T., \& Shull, J.M. 2002, ApJ 565, 720

\bref
Petitjean, P., Webb, J.K., Rauch, M., Carswell, R. F., \& Lanzetta, K. 1993,
MNRAS 262, 499

\bref
Putman, M.E., Bland-Hawthorn, J., Veilleux, S., et al.\,2003, ApJ 597, 948

\bref
Rand, R.J. 1996, ApJ 462, 712

\bref
Rhee, M.-H., \& van Albada, T.S. 1996, A\&AS 115, 407

\bref
Richter, P., de Boer, K.S., Widmann, H., et al. 1999,
{\it Nature} 402, 386

\bref
Richter, P., Savage, B.D., Wakker, B.P., Sembach, K.R., Kalberla, P.M.W.
2001a, ApJ 549, 281

\bref
Richter, P., Sembach, K.R., Wakker, B.P., et al. 2001b,
ApJ 559, 318

\bref
Richter P.,  Sembach K.R., Wakker B.P.,
\& Savage B.D. 2001c, ApJ 562, L181

\bref
Richter, P., Wakker, B.P., Savage, B.D., Sembach, K.R. 2003a, ApJ 586, 230

\bref
Richter, P., Sembach, K.R., Howk, J.C. 2003b, A\&A 405, 1013

\bref
Richter, P., \& de Boer, K.S. 2004, 
in High-velocity Clouds, ed. H. van Woerden et al.,
ASSL, Vol.\,312 (Kluwer), ISBN 1402025785, 183

\bref
Richter, P., Savage, B.D., Tripp, T.M., \& Sembach, K.R. 2004, ApJS, 153, 165

\bref
Richter, P., Westmeier, T., \& Br\"uns, C. 2005, A\&A 442, L49

\bref
Richter, P., Savage, B.D., Sembach, K.R., \& Tripp, T.M. 2006a, 
A\&A 445, 827

\bref
Richter, P., Fang, T., \& Bryan, G,L. 2006b, A\&A, in press; astro-ph 0511609

\bref
Rossa, J., \& Dettmar, R.-J. 2003, A\&A 406, 493

\bref
Savage, B.D., \& de Boer, K.S. 1979, ApJ 230, L77

\bref
Savage, B.D., \& de Boer, K.S. 1981, ApJ 244, 768

\bref
Savage, B.D., Sembach, K.R., \& Lu, L. 1997, AJ 113, 2158

\bref
Savage, B.D., Sembach, K.R., Tripp, T.M., \& Richter, P. 2002, ApJ 564, 631

\bref
Savage, B.D., Sembach, K.R., Wakker, B.P., et al. 2003, ApJS 146, 125

\bref
Savage, B.D., Lehner, N., Wakker, B.P., Sembach, K.R, \& Tripp, T.M. 2005, ApJ 626, 776

\bref
Sembach, K.R., Savage, B.D., \& Massa, D. 1991, ApJ 372, 81

\bref
Sembach, K.R., \& Savage, B.D. 1992, ApJS 83, 147

\bref
Sembach K.R., Howk J.C., Savage B.D., \& Shull J.M. 2001, AJ 121, 992

\bref
Sembach K.R., Wakker, B.P., Savage, B.D., et al. 2003, ApJS 146, 165

\bref
Sembach, K.R., Wakker, B.P., Tripp, T.M., et al. 2004a, ApJS 150, 387

\bref
Sembach, K.R., Tripp, T.M., Savage, B.D., \& Richter, P. 2004b, ApJS 155, 351

\bref
Shapiro, P.R., \& Field, G.B. 1976, ApJ 205, 762

\bref
Sirko, E., Goodman, J., Knapp, G.R., et al.\,2004, AJ 127, 899

\bref
Spitzer, L. 1956, ApJ 124, 20

\bref
Steidel, C.C, Dickinson, M., \& Persson, E. 1994, ApJ 437, L75

\bref
Strickland, D.K., Heckman, T.M., Colbert, E.J.M., Hoopes, C.G., Weaver, K.A. 2004,
ApJ 606, 829

\bref
Thilker, D.A., Braun, R., Walterbos, R.A.M., et al. 2004, ApJ 601, L39

\bref
Thom, C., Putman, M.E., \& Gibson, B.K. 2005, ApJL, in press

\bref
Tripp, T.M., Savage, B.D., \& Jenkins, E.B. 2000, ApJ 534, L1

\bref
Tripp, T.M., Wakker, B.P., Jenkins, E.B., et al. 2003, AJ 125, 3122

\bref
Valageas, P., Schaeffer, R., \& Silk J. 2002, A\&A 388, 741

\bref
van Woerden, H., Schwarz, U. J., Peletier, R. F., Wakker, B. P., \& Kalberla, P. M. W.
1999, {\it Nature} 400, 138

\bref
Wannier, P., \& Wrixon, G.T. 1972, ApJ 173, L119

\bref
Weiner, B.J., Vogel, S.N., Williams, T.B. 2001, in ASP Conf.\,Ser.\,240,
Gas and Galaxy Evolution, ed. J.E. Hibbard, M.Rupen \& J. van Gorkom, 515

\bref
Wakker, B.P., Howk, J.C., Savage, B.D., et al. 1999, Nature, 402, 388

\bref
Wakker, B.P. 2001, ApJS 136, 463

\bref
Wakker, B.P., Savage, B.D., Sembach, K.R., et al. 2003, ApJS 146, 1

\bref
Wakker, B.P. 2004, in High-velocity Clouds, ed. H. van Woerden et al.,
ASSL, Vol.\,312 (Kluwer), ISBN 1402025785, 25

\bref
Wakker, B.P., \& Richter, P. 2004, {\it Scientific American} 290, 28

\bref
Wang, Q.D., Immler, S., Walterbos, R., Lauroesch, J.T.; Breitschwerdt, D. 2001, ApJ 555, L99

\bref
Wang, Q.D. 2005, in Extra-planar gas, ed. R. Braun, ASP Conference Series,  Vol.\,331, 329

\bref
Westmeier, T., Br\"uns, C., \& Kerp, J. 2005a, 
in Extra-planar gas, ed. R. Braun, ASP Conference Series,  Vol.\,331, 105

\bref
Westmeier, T., Braun, R., \& Thilker, D. 2005b, A\&A 436, 101

\bref
Weymann, R., Jannuzi, B.T., Lu, L., et al. 1998, ApJ 506, 1

\bref
Widmann, H., de Boer, K. S., Richter, P., et al. 1998, A\&A 338, L1

\bref
Wolfire, M.G., McKee, C.F., Hollenbach, D., \& Thielens, A.G.G.M. 1995, ApJ, 453, 673

}

\vfill

\end{document}